\documentclass[10pt]{article}
\usepackage{amssymb}
\usepackage{verbatim}
\makeatletter
\@addtoreset{equation}{section}

\def\bbN{{\mathbb N}}
\def\ds{\displaystyle}

\def\un{\underline}

\def\vrul{\rule[20pt]{0pt}{0pt}}
\newtheorem{theorem}{Theorem}[section]
\newtheorem{examp}{Example}[section]
\newtheorem{coroll}{Corollary}[section]
\newtheorem{examps}{Examples}[section]

\newtheorem{lemma}{Lemma}[section]
\newtheorem{remark}{Remark}[section]

\newtheorem{remarks}[remark]{Remarks}
\newtheorem{proposition}{Proposition}[section]
\newtheorem{definition}{Definition}[section]

\def\le{\left}
\def\m{\mathop}

\def\Bmat{\mathop{\mathbb  B}}

\def\Amat{\mathop{\mathbb A}}
\def\ri{\right}
\def\br{\begin{remark}\rm\small}
\def\1{{\bf 1}}
\def\er{\end{remark}}
\def\bt{\begin{theorem}\rm}
\def\et{\end{theorem}}
\def\bc{\begin{coroll}\rm}
\def\ec{\end{coroll}}
\def\brs{\begin{remarks}.\\ \rm\small\begin{enumerate}}
\def\ers{\end{enumerate}\end{remarks}}
\def\bx{\begin{examp}\small}
\def\ex{\end{examp}}
\def\bl{\begin{lemma}\small}
\def\el{\end{lemma}}
\def\bxs{\begin{examps}. \rm\begin{enumerate}}
\def\exs{\end{enumerate}\end{examps}}
\def\bd{\begin{definition}}
\def\ed{\end{definition}}
\def\bp{\begin{proposition}\rm}
\def\ep{\end{proposition}}
\def\be{\begin{equation}}
\def\ee{\end{equation}}

\def\bea{\begin{eqnarray}}
\def\eea{\end{eqnarray}}
\def\beas{\begin{eqnarray*}}
\def\eeas{\end{eqnarray*}}
\def \pa{\partial}
\def\C{{\mathbb C}}
\def\A{\mathop{\mathbf a}}
\def\B{\mathop{\mathbf b}}

\def\N{{\mathbb N}}

\def\a{{\alpha}}
\def\b{{\beta}}
\def\ra{\rightarrow}
\def\DD{{\mathcal D}}
\def\MC{{\mathcal M}}
\def\MM{{\mathbf M}}
\def\LL{{\mathbf L}}
\def\di{{\partial}}
\def\KK{{\mathbf K}}

\def\wtKKh{{\hat{\widetilde{\mathbf K}}}}
\def\wtKK{{\widetilde{\mathbf K}}}
\def\wtK{{\widetilde K}}
\def\bfu{{\mathbf u}}
\def\bfP{{\mathbf P}}

\def\wtf{{\widetilde f}}
\def\wtg{{\widetilde g}}
\def\bff{{\mathbf f}}
\def\bfg{{\mathbf g}}
\def\bfwtf{{\widetilde{\mathbf f}}}
\def\bfwtg{{\widetilde{\mathbf g}}}
\date{}
\begin{document}
\begin{titlepage}
\begin{flushright}
CRM-2749 (2001)\\
\hfill Saclay-T01/047\\
nlin.SI/0108049
\end{flushright}
\vspace{0.2cm}
\begin{center}
\begin{Large}
\textbf{Duality, Biorthogonal Polynomials and Multi--Matrix Models}\footnote{
Work supported in part by the Natural Sciences and Engineering Research Council
of Canada (NSERC) and the Fonds FCAR du Qu\'ebec.}
\end{Large}\\
\vspace{1.0cm}
\begin{large} {M.
Bertola}$^{\dagger\ddagger}$\footnote{bertola@crm.umontreal.ca}, 
 { B. Eynard}$^{\dagger
\star}$\footnote{eynard@spht.saclay.cea.fr}
 and {J. Harnad}$^{\dagger \ddagger}$\footnote{harnad@crm.umontreal.ca}
\end{large}
\\
\bigskip
\begin{small}
$^{\dagger}$ {\em Centre de recherches math\'ematiques,
Universit\'e de Montr\'eal\\ C.~P.~6128, succ. centre ville, Montr\'eal,
Qu\'ebec, Canada H3C 3J7} \\
\smallskip
$^{\ddagger}$ {\em Department of Mathematics and
Statistics, Concordia University\\ 7141 Sherbrooke W., Montr\'eal, Qu\'ebec,
Canada H4B 1R6} \\ 
\smallskip
$^{\star}$ {\em Service de Physique Th\'eorique, CEA/Saclay \\ Orme des
Merisiers F-91191 Gif-sur-Yvette Cedex, FRANCE } \\
\end{small}
\bigskip
\bigskip
{\bf Abstract}
\end{center}
\begin{small}

The statistical distribution of eigenvalues of pairs of coupled
random matrices can be expressed in terms of integral kernels having a
generalized  Christoffel--Darboux form constructed from sequences of
biorthogonal  polynomials.  For measures involving exponentials of a pair of
polynomials  $V_1$, $V_2$ in two different variables, these kernels may be
expressed in terms of finite dimensional ``windows'' spanned by finite
subsequences having length equal to the degree of one or the other of the
polynomials $V_1$, $V_2$. The vectors formed by such subsequences satisfy
``dual pairs'' of first  order systems of linear differential equations with
polynomial coefficients, having rank equal to one of the degrees of $V_1$ or
$V_2$ and degree equal to the other. They also satisfy recursion relations
connecting the consecutive windows, and deformation equations,  determining how
they change under variations in the coefficients of the polynomials $V_1$ and
$V_2$. Viewed as overdetermined systems of linear
difference-differential-deformation equations, these are shown to be 
compatible, and hence to admit simultaneous fundamental systems of solutions.
The main result is the demonstration of a spectral duality property; namely,
that the  spectral curves defined by the characteristic equations of the pair
of matrices defining the dual differential systems are equal upon interchange
of eigenvalue and polynomial parameters.
\end{small}

\end{titlepage}
%

\section{Introduction}

\subsection{Random matrices}

\subsubsection{Background and motivation}

Random matrices \cite{Mehta, BI} play an important r\^ole in many areas of 
physics.  They were first introduced by Wigner \cite{Wigner} in the  context
of the spectra of large nuclei, and the theory was greatly developed in
pioneering work of Mehta, Gaudin \cite{MehtaGaudin, Gaudin} and Dyson
\cite{Dyson1, Dyson2}. It has found subsequent applications in  solid state
physics \cite{Guhr} (e.g., conduction in mesoscopic devices,  quantum chaos and,
lately, crystal growth\cite{spohn}), in particle  physics \cite{Verbaarshot},
$2d$-quantum gravity and string theory \cite{Matrixsurf, ZJDFG, DVV}. The
reason for the success and large range of applications of random matrices is,
due, in part, to their universality property; when the size of the  matrices
$N$ becomes large, the statistics of the eigenvalues tend to be  independent of
the model, and determined only by its symmetries and the spectral region 
considered, relative to critical points and edges in the spectral density.
Matrix integrals are also known to give special realizations of $KP$, Toda and
isomonodromic $\tau$-functions, and thus have a close relationship to
integrable  systems \cite{DouglasKdV,  KostovADE, TW1, TW2, AvM1, AvM2, HTW,
BI}.

Random matrices also have important applications in pure mathematics, for 
example, in the statistical distribution  of the zeros  of $\zeta$-functions
\cite{Odl, RS, KS}. They are also related to other  statistical problems such as
random  word growth and the lengths of nondecreasing subequences of random
sequences
\cite{BDJ, Joh}. 

The model we shall consider here is called the ``2-matrix model''
\cite{ItzyksonZuber, Mehtaorthpol, KazakoVDK, MehtaShukla, ZJDFG, eynard,
 eynardmehta}. This involves an ensemble consisting of pairs of 
$N\times N$ hermitian matrices $M_1$ and $M_2$, with a $U(N)$ invariant
probability measure of the  form: 
\be
{1 \over \tau_N}d\mu(M_1,M_2):= {1\over \tau_N} \exp{{\rm tr}\, 
\left( -V_1(M_1) - V_2(M_2) + M_1 M_2 \right)} dM_1 dM_2  \ , 
\label{twomatrixmeas}
\ee
where $dM_1 dM_2$ is the standard Lebesgue measure for pairs of Hermitian 
matrices, $V_1$ and $V_2$ are polynomials of degrees $d_1+1$, $d_2+1$ 
respectively, called the potentials, with coeffficients viewed as
deformation parameters, and the normalization factor (partition function) is 
\be
\label{deftau}
\tau_N = \int_{M_1}\int_{M_2} d\mu \ ,
\ee
which is known to be a KP $\tau$-function in each set of deformation
parameters, as well as providing solutions to the two-Toda equations
\cite{UT, AvM1, AvM2}.

This model was introduced in \cite{ItzyksonZuber, Mehtaorthpol}
as a toy  model for quantum gravity and string theory. The main interest was
in a special ``double scaling'' limit, where $N\to\infty$  and the potentials
$V_1$ and $V_2$ are fine-tuned to critical potentials. The asymptotic
behaviour in  such limits is  related to finite dimensional irreducible
representations  of the 2D-conformal group \cite{Matrixtwoconform, Matrixconform, KazakoVDK}.
The best known example is when $V_1$ and $V_2$ are cubic polynomials,  tuned
to  their critical values, which reproduces the critical behaviour of the Ising
 model on a random surface \cite{Kazakov, KazakovIsing}. It is important to 
note that the 2-matrix model contains more critical points than a 1-matrix
model; for instance, the 1-matrix-model cannot have an Ising  transition
\cite{Matrixtwoconform}.

String theorists have also introduced a generalization, known as the 
``multi-matrix model'' \cite{DouglasKdV,  eynard, eynardchain, eynardmehta},
where one has a set of $m\geq 2$ matrices   ($N\times N$ hermitian) coupled
together in a chain, with a measure of the form
\be
{1\over \tau_N}d\mu(M_1, \dots M_n)=
{1\over \tau_N} \exp{{\rm tr}\, \left( -\sum_{j=1}^mV_j(M_j)
 + \sum_{j=1}^{m-1}M_i M_{i+1}\right)} \prod_{i=1}^m dM_i \ ,
\ee
and the $V_j$'s are again polynomials in their arguments.
This model has the same universal behaviour as the 2-matrix model and, in
some sense, does not seem to contain any more information.  
Throughout the main body of this work, we will concentrate on the
2-matrix-model, for which the statistics of the eigenvalues can be calculated
using biorthogonal  polynomials \cite{Mehtaorthpol, Mehta, eynardmehta,
eynardchain, AvM1, AvM2}. In the appendix, it will be explained how to extend
all the results in the present work from the 2-matrix model to the differential
systems associated with this
 multi-matrix model.

\subsubsection{Relation to biorthogonal polynomials}

\par
By biorthogonal polynomials, we mean two sequences of monic polynomials
\be
\pi_n(x) = x^n + \cdots , \qquad \sigma_n(y)=y^n + \cdots, \qquad n=0,1,\dots
\ee
which are orthogonal with respect to a coupled measure on the product space:
\be
\int\int {\rm d}x\, {\rm d}y \,\, \pi_n(x)\sigma_m(y) {\rm
e}^{-V_1(x)-V_2(y) +xy} = h_n\delta_{mn} ,
\ee 
where $V_1(x)$ and $V_2(y)$ are polynomials chosen to be the same as those
appearing in the 2-matrix model measure (\ref{twomatrixmeas}), and a suitable 
contour is chosen to make the integrals convergent.  The orthogonality 
relations determine the two families.  
  Once the biorthogonal polynomials are known, they may be used to compute 
four different kernels:

\bea
&&{\m{K}^N}_{12}(x,y) =  \sum_{n=0}^{N-1} {1\over h_n} \pi_n(x) 
\sigma_n(y) {\rm  e}^{-V_1(x)} {\rm  e}^{-V_2(y)}  \ ,   \qquad
{\m{K}^N}_{11}(x,x') = \int {{\rm d}y \,\, {\m{K}^N}_{12}(x,y)\, {\rm  e}^{x'
y}}, \\ &&{\m{K}^N}_{22}(y',y) = \int {{\rm d}x \,\, {\m{K}^N}_{12}(x,y)\,
{\rm  e}^{x y'}}  \ ,  \qquad   {\m{K}^N}_{21}(y',x') = \int \int {{\rm d}x
\,{\rm d}y  {\m{K}^N}_{12}(x,y)\, {\rm  e}^{xy'}{\rm  e}^{x'y}}  \ . 
\label{Ker12} \eea

All the statistical properties of the spectra of the 2-matrix ensemble may
then be expressed in terms of these kernels \cite{eynardmehta} and the
corresponding Fredholm integral operators $\ds{{\m{\KK}^N}_{ij}}, \ i,j=1,2$.
For instance the density of eigenvalues of the first matrix is:
\be 
{\m{\rho}^N}_1(x) = {1\over N}\, {\m{K}^N}_{11}(x,x) \ ,
\ee
the correlation function of two eigenvalues of the first matrix is:
\be 
{\m{\rho}^N}_{11}(x,x') = {1\over N^2}\left({\m{K}^N}_{11}(x,x) {\m{K}^N}_{11}(x',x') 
- {\m{K}^N}_{11}(x,x') {\m{K}^N}_{11}(x',x)\right) \ ,
\ee
and the correlation function of two eigenvalues, one of the first matrix and one
of the second is:
\be
{\m{\rho}^N}_{12}(x,y) = {1\over N^2}
\left({\m{K}^N}_{11}(x,x) {\m{K}^N}_{22}(y,y) - 
{\m{K}^N}_{12}(x,y) ({\m{K}^N}_{21}(y,x) - e^{x y})\right) \ . 
\ee
Any other correlation function of $m$ eigenvalues can similarly be written 
as a determinant involving these four kernels only.

The spacing distributions (the probability that two neighbouring eigenvalues 
are at some given distance) can be computed as Fredholm determinants. For
example, the probability that some subset $J$ of the real axis contains no 
eigenvalue of the first matrix is the Fredholm determinant:
\be 
p^{N,1}_J = \det\left( \1 - {\m{\KK}^N}_{11} \circ \chi_J \right) ,
\ee
where $\chi_J$ is the characteristic function of the set $J$.

An important feature in the study of the $N\ra \infty$ limit is that the
kernels $K_{ij}$ may be expressed \cite{eynardchain} in terms of sums involving
only a fixed  number of terms (either $d_1+1$ or $d_2+1$), independently of
$N$, as a  consequence of a ``generalized Christoffel--Darboux'' formula
\cite{Szego, Userkesm} following from the recursion relations
satisfied by the biorthogonal polynomials. This allows one, in the $N\ra
\infty$ limit, with suitable scaling in the spectral variables, depending on
the region considered, to treat $N$ as just a  parameter.

\bigskip

\subsection{Duality}
\subsubsection{Dual isomonodromic deformations}

The notion of {\it duality} arises in a number of contexts, both in relation 
to isospectral flows \cite{AHH} and isomonodromic \cite{H1, H2, HI}
deformations. What is meant here by ``duality'' in the case of isomonodromic
deformations is the existence of a pair of parametric families of meromorphic
covariant derivative operators on the Riemann sphere
\bea
\DD_1:= {\di \over \di x} + \LL(x,\bfu) \ , \qquad
\DD_2:= {\di \over \di y} + \MM(y,\bfu) \ ,\label{DD12}
\eea 
where $\LL(x,\bfu)$ and $\MM(y,\bfu)$ are, respectively, $l\times l$ and
$m\times m$  matrices that are rational functions of the complex variables 
$x\in \bfP^1$ and  $y\in \bfP^1$, with pole divisors of fixed 
degrees, depending smoothly on a set of deformation parameters 
$\bfu= (u_1, u_2, \dots)$ in such a way that: 
\newline \noindent 
1)
The matrices  $\LL(x,\bfu)$ and $\MM(y,\bfu)$ are obtained from the integral
curves of a set of commuting (in general, nonautonomous) vector fields defined
on a phase space $\MC$ by composition with a prescribed pair of maps (possibly
depending explicitly  on the deformation parameters) from $\MC$ to the spaces 
of rational, $l\times l$  or  $m\times m$ matrix valued rational functions of 
the spectral parameter $x$ or $y$, respectively, with pole divisors of fixed 
degree.  
\newline \noindent
2) The generalized monodromy data of both the operators $\DD_1$ and $\DD_2$
are invariant under the $\bfu$-deformations. (This includes the monodromy
representation of the fundamental group of the punctured Riemann sphere
obtained by removing the locus of poles and, in the case of non-Fuchsian
systems,  the Stokes matrices and connection matrices \cite{JMU, JM}.)  
\newline \noindent
3) The spectral curves determined by the characteristic equations:
\bea
\det(\LL(x, \bfu) - y\1 )=0  \ , \qquad
\det(\MM(y, \bfu) - x\1 )=0   \label{MMLL}
\eea
are biholomorphically equivalent.
\newline \noindent
(A similar definition can be given for the case of dual isospectral
flows of matrices $\LL(x, \bfu)$ and $\MM(y, \bfu)$.) 

   Such ``dual pairs'' of isomonodromic families occur in many applications.
They are  related to the solution of ``dual'' matrix Riemann-Hilbert (RH)
problems \cite{IIKS, HI, H2} which, in certain cases, are equivalent to determining
 the resolvents of  a special class of ``integrable'' Fredholm integral 
operators $\KK$, $\wtKK$, with kernels of the form:
\bea
&& K(x,x')  =  \sum_{i=1}^l {f_i(x) g_i(x') \over x- x'}\ , \label{intK}\\
&& \wtK(y,y')  =  \sum_{a=1}^m {\wtf_a(y) \wtg_a(y') \over y -
y'}\ .\label{intwtK}
\eea
Here, the vector valued functions
\bea
&&\bff(x,\bfu) =(f_1(x,\bfu), \dots f_l(x,\bfu))  \ ; \qquad
\bfg(x,\bfu) =(g_1(x,\bfu), \dots g_l(x,\bfu))   \\
&&\bfwtf(y,\bfu) =(\wtf_1(y,\bfu), \dots \wtf_m(y,\bfu))  \ ; \qquad
\bfwtg(y,\bfu) =(\wtg_1(y,\bfu), \dots \wtg_m(y,\bfu)) \ . \label{fgdef}
\eea
depend on the spectral variables $x$ and $y$ as well as on some, but not
necessarily all the deformation parameters $(u_1, u_2, \dots)$. They
satisfy overdetermined, compatible differential systems in these variables
which imply the invariance of the monodromy of  associated ``vacuum''
isomonodromic families of covariant derivative operators
\bea
\DD_{0,1}:= {\di \over \di x} + \LL_0(x,\bfu) \ , \qquad
\DD_{0,2}:= {\di \over \di y} + \MM_0(y,\bfu) \ .\label{DD12vac}
\eea 
 The  dual kernels $K(x,x')$ and $\wtK(y,y')$ are related to each
other by applying partial integral transforms with respect to one of the
two spectral variables $(x,y)$ (e.g., Fourier-Laplace transforms) to an
integral operator on the product space $\bfP^1\times \bfP^1$. Application of
the Riemann-Hilbert dressing method for suitably chosen ``dual'' sets
of discontinuity data then gives rise to the ``dressed'' families (\ref{DD12}),
which have a similar relation to the resolvent kernels of the two operators.
(Additional deformation parameter dependence may enter, besides
that contained in the vacuum equations, characterizing the support of these
operators.)  The Fredholm determinants $\det(\1 -\KK )$ and
$\det(\1 - \wtKKh)$ may then be shown, through deformation formulae, to
coincide with the corresponding isomonodromic tau functions, and with
each other (see e.g. \cite{HI, H2}).

   Such systems arise naturally, as discussed above, in the study of the
spectral statistics of random matrix ensembles, both in the finite case, and
in suitable infinite limits. An example of such dual pairs is given by the
class of exponential kernels in which:
\bea
&&f_i(x)= (-1)^i e^{u_i x}  \ , \qquad
g_i(x)= e^{-u_i x} \ ,  \\
&&\wtf_a(x)= (-1)^a e^{v_a x}  \ , \qquad
\wtg_a(x)= e^{-v_a x}  \ ,  \label{expfgdef}
\eea
For the case $l=m=2$, these include the sine kernel
\be
K_S(x,x')= {\sin(u (x-x'))\over  (x-x')},
\qquad \wtK_S(y,y')= {\sin(v (y-y'))\over  (y-y')}  \label{sineker}
\ee
governing the spectral statistics in the scaling limit of the GUE in the bulk
region \cite{Mehta}.

   Other examples include the various Painlev\'e equations $P_{II}$, $P_{IV}$,
$P_{V}$, $P_{VI}$ \cite{H1, HTW, Du} which each possess ``dual'' isomonodromic
representations. A last class of examples, unrelated to random matrices, but
including a special case of $P_{VI}$, consists of the isomonodromic
representations of the WDVV equations of topological 2D gravity entering in 
the theory of Frobenius manifolds \cite{Du}. These possess both Fuchsian
representations with $n$ finite poles with residues of rank $1$, and 
non-Fuchsian $n\times n$ representations, having a single irregular singular 
point of Poincar\'e index $1$ at $\infty$.

\subsubsection{Duality in the large $N$ Limit}

It is proved in \cite {PZJ}, under a suitable large $N$ assumption, and
conjectured for other cases \cite{PZJ,  eynardchain} that in a particular
large $N$ limit (where the coefficients $\{u_K, \ v_J\}$ scale as $N$, and the
supports of ${\ds \rho_1(x):={\rm lim}_{N\to\infty}{\m \rho^N}_1(x)} $ and
${\ds\rho_2(y):={\rm lim}_{N\to\infty} {\m \rho^N}_2(y)}$ are connected
intervals 
$[a_1,b_1]$,
$[a_2,b_2]$ respectively), the following functions:
\be
f(x) := {1\over N}V'_1(x) - \int_{a_1}^{b_1} {\rho_1(x')\over x-x'} {\rm d} x'
\ , \quad 
g(y) := {1\over N}V'_2(y) - \int_{a_2}^{b_2} {\rho_2(y')\over y-y'}
{\rm d} y' \ , \quad
\ee
 are inverses of each other. In other words, if $y=f(x)$ then $g(y)=x$.
One can see that the functions $f(x)$ and $g(y)$ are related to the 
eigenvalues of the operators which implement the derivative with respect to $x$
and $y$ for the biorthogonal polynomials. The spectral duality theorems
\ref{maintheorem} and \ref{system_duality} which we present in this work are
a more precise statement related to this conjecture, with a rigorous proof
valid for all $N$.

\subsection{Outline of the article}

\subsubsection{Biorthogonal polynomials and differential systems}

\indent
In Section 2, we consider the normalized quasi-polynomials
\be
 \psi_n(x) = \frac 1{\sqrt{h_n}} \pi_n(x){\rm e}^{-V_1(x)} \ , \qquad
\phi_n(y) = \frac 1{\sqrt{h_n}} \sigma_n(y) {\rm e}^{-V_2(y)}\ , 
n=0, \dots \infty  \ .
\ee
Viewing these as the components of a pair of
column vectors 
\be
\ds{\m{\Psi}_\infty}= (\psi_0, \psi_1, \dots \psi_n, \dots )^t \quad
{\rm and } \quad
 \ds{\m{\Phi}_\infty}=(\phi_0, \phi_1, \dots \phi_n, \dots )^t \ ,
\ee
 we obtain 
a pair of semi-infinite matrices $Q$ and $P$ that implement
multiplication of $\ds{\m{\Psi}_\infty}$ by $x$ and derivation with respect to
$x$, respectively. Equivalently, we obtain the transposes $Q^t$ and
$P^t$ by applying $-{d \over dy}$ or multiplication by $-y$ to 
$\ds{\m{\Phi}_\infty}$. By construction, these satisfy the Heisenberg 
commutation relations
\be
[P, \ Q] = \1 \ ,
\ee
and, as shown in Section 2, they are finite band matrices;
$Q$ has nonvanishing elements only along diagonals that range from $1$ above
the principal diagonal to $d_2$ below it, and $P$ has nonvanishing elements
only along the diagonals from $1$ below the principal to $d_1$ above it, where
$d_1+1$ and $d_2+1$ are the degrees of the polynomials $V_1(x)$ and $V_2(y)$,
respectively. 

   The first result (Prop.\ref{generalDC}) following from the finite recursion 
relations satisfied by the  quasi-polynomials $\{\psi_n(x)\}$ and 
$\{\phi_n(y)\}$ is a set of ``generalized Christoffel--Darboux relations 
\cite{Userkesm, eynardchain}, which imply that the kernels
$\ds{\m{K}^N}_{11}(x,x')$ and 
$\ds{\m{K}^N}_{22}(y',y)$ may be expressed as: 
\be
{\m{K}^N}_{11}(x,x')  = 
-  {\ds{\le(\m{\un\Phi}^{N\!-\!1} (x'),\Amat^N  \m{\Psi}_{N} (x)\ri)}
\over x-x'} \ ,
\qquad
  {\m{K}^N}_{22}(y',y) =
 {\ds{\le(\m{\un\Psi}^{N\!-\!1} (y'), \Bmat^N  \m{\Phi}_{N} (y)\ri)} 
\over y'-y}  \ ,
\ee
where  $\ds{\m{\Psi}_{N} (x)}$ and $\ds{\m{\Phi}_{N} (y)}$ are the $d_2+1$ and 
$d_1+1$ dimensional column vectors with components
$[\psi_{N\!-\!d_2},\dots,\psi_{N}]$ and $[\phi_{N\!-\!d_2},\dots,\phi_{N}]$,
respectively, and $\ds{\m{\un\Psi}^{N-1} (y)}$ and $\ds{\m{\un\Phi}^{N-1} (x)}$ 
are the $d_2+1$  and $d_1+1$ component row vectors with components
$[\un\psi_{N\!-\! 1},\dots,\un\psi_{N\!+\!d_2\!-\!1}]$ and
$[\un\phi_{N\!-\! 1},\dots,\un\phi_{N\!+d_1\!-\! 1}]$, respectively, where the
underbarred quantities $\{\un\psi_n(y)\}$  and $\{\un\phi_n(y)\}$ designate
the Fourier-Laplace transforms of the quasi-polynomials $\{\psi_n(y)\}$ and
$\{\phi_n(y)\}$. The matrices $\ds{\Amat^N} $ and $\ds{\Bmat^N}$ are, 
essentially, the nonvanishing parts of the matrices obtained by commuting 
$Q$ and $P$,
respectively, with the projectors to the appropriate finite-dimensional
subspace.  A similar ``differential'' form of the generalized
Christoffel--Darboux relations holds, following from applying the derivations
$\partial_x + \partial_{x'}$ and  $\partial_y + \partial_{y'}$ to the kernels
$\ds{\m{K}^N}_{11}(x,x')$ and $\ds{\m{K}^N}_{22}(y',y)$.

  The recursion relations satisfied by the quasi-polynomials  
$\{\psi_n(x)\}_{n=0\dots \infty}$ and $\{\phi_n(y)\}_{n=0\dots \infty}$ may be
conveniently expressed (Lemma \ref{abNrecursions}) as:
$$
\A_N\m{\Psi}_N(x) =
\m{\Psi}_{N\!+\!1}(x)\ , \qquad
\B_N\m{\Phi}_N(y) =
\m{\Phi}_{N\!+\!1}(y)  \ ,
$$
where the ``ladder'' matrices $\ds{\A_N}$ and $\ds{\B_N}$ are linear in $x$ 
and in $y$, respectively, and are formed from the rows of $Q$ and the columns 
of $P$ (see eqs. (\ref{aNdef}, \ref{bNdef}) for their exact definitions).

The vectors $\ds{\m{\Psi}_{N} (x)}$ and $\ds{\m{\Phi}_{N}(y)}$ also satisfy 
the following differential equations (Lemma \ref{D12})
\be
\frac \pa{\pa x} \m{\Psi}_N = -\m{D_1}^N (x)
 \m{\Psi}_N  \ ,    \qquad
\frac \pa{\pa y} \m{\Phi}_N = -\m{D_2}^N (y)
 \m{\Phi}_N \ , 
\ee
where the matrices $\ds{\m{D_1}^N(x)}$ and $\ds{\m{ D_2}^N(y)}$ are, 
respectively, of size $(d_2+1)\times (d_2+1)$ and  $(d_1+1)\times (d_1+1)$, 
with entries that are polynomials in the indicated variables of 
degrees $d_1$ and $d_2$, respectively (and also
polynomials in the matrix entries of $Q$ and $P$). Furthermore, if $\{u_K
\}_{K=1\dots d_1+1}$ and  $\{v_J\}_{J=1\dots d_2+1}$  are the coefficients
of the polynomials $V'_1(x)$ and $V'_2(y)$, respectively, and these
are varied smoothly, the effect of such deformations is given by the
following system of PDE's (Lemma \ref{uvPsiNPhiNdefs})
\bea
 \frac \pa{\pa u_K} \m{\Psi}_N  &=& {\m{\bf U}^{N,\Psi}}_{K}\m{\Psi}_N  \ , 
\qquad 
 \frac \pa{ \pa v_J} \m{\Psi}_N  = -{\m{\bf V}^{N,\Psi}}_{J}\m{\Psi}_N   \ , 
\cr
 \frac \pa{\pa u_K} \m{\Phi}_N &=& {\m{\bf U}^{N,\Phi}}_{K}\m{\Phi}_N \ , 
\qquad 
 \frac \pa{\pa v_J} \m{\Phi}_N  = -{\m{\bf V}^{N,\Phi}}_{J}\m{\Phi}_N  \ ,
\eea
where the matrices $\ds{\m{\bf U}^{N,\Psi}}_{K}(x)$, $\ds{\m{\bf
V}^{N,\Psi}}_{J}(x)$,$\ds{\m{\bf U}^{N,\Phi}}_{K}(y)$ and $\ds{\m{\bf
V}^{N,\Phi}}_{J}(y)$ are again polynomials in the indicated variables and in
the matrix entries of $Q$ and $P$.

\subsubsection{Compatibility} 

\indent
   So far, these statements are just a re-writing of the infinite series of
recursion relations, differential equations and deformation equations satisfied
by the functions $\{\psi_n(x)\}_{n=0\dots \infty}$ and
$\{\phi_n(y)\}_{n=0\dots \infty}$, projected onto the finite ``windows''
represented by the vectors $\ds{\m{\Psi}_N}$ and $\ds{\m{\Phi}_N}$. However, 
we may now view these equations as defining an overdetermined system of
finite difference-differential-deformation equations for vector
functions of the variable $\{N, x,y, u_K, v_J\}$, and ask whether, as such,
these systems are {\it compatible}; i.e., whether they admit a {\it basis} of
simultaneous linearly independent solutions. The affirmative answer to this
question is provided in  Section 3 by Prop. \ref{dx_du_compatibility}, which
states that sequences of invertible $(d_2+1) \times (d_2+1)$ and  $(d_1+1)
\times (d_1+1)$ matrices $\ds{\m{\bf \Psi}_{N}(x)}$ and 
$\ds{\m{\bf \Phi}_{N}(y)}$ exist (fundamental solutions),  for which all the 
column vectors satisfy the above difference-differential-deformation equations 
simultaneously.

The compatibility of the deformation equations and finite difference equations
with the $x$ and $y$ differential equations imply, in particular, that the
(generalized)  monodromy of the polynomial covariant derivative operators
\be
{\pa \over \pa x} + \m{D_1}^N (x) \ , \qquad 
{\pa \over \pa y} + \m{D_2}^N (y) 
\ee
is invariant under both the  $\{u_K, v_J\}$ deformations and the shifts in $N$.
A similar statement can be made of the corresponding operators
\be
{\pa \over \pa x} - \m{\un D_1}^N (x) \ , \qquad 
{\pa \over \pa y} - \m{\un D_2}^N (y) 
\ee
annihilating the Fourier-Laplace transformed vectors 
$\ds{\m{\un\Phi}^{N-1} (x)}$ and $\ds{\m{\un\Psi}^{N-1} (y)}$.

\subsubsection{Spectral duality}

\indent
   A result that is not at all obvious from the above discussion is proved in
Prop. \ref{accordion}; namely, that the pairs of matrices $\ds{(\m{D_1}^N (x),
\m{\un D_2}^N (y) )}$ and $\ds{(\m{D_2}^N (y), \m{\un D_1}^N (x))}$ have the 
same {\it spectral curves}. More specifically, we have the following 
equalities between their characteristic equations, which actually are 
identities 
\bea
&& u_{d_1\!+\! 1}\det\le[x\1_{d_1+1 }-{\m{D}^N}_2 (y) \ri]  =
v_{d_2\!+\! 1}\det\le[y\1_{d_2+1} - {\m{\un D}^N}_1 (x)\ri]
  \\
&& u_{d_1\!+\! 1} \det\le[x\1_{d_1+1 }-{\m{\un D}^N}_2(y) \ri]  =
v_{d_2\!+\! 1} \det\le[y\1_{d_2+1} -  {\m{D}^N}_1 (x)\ri]\ .
\eea
(Note that the two integers $d_1 +1$ and $d_2+1$ which determine, in one case,
the dimension of the matrix, in the other, its degree as a polynomial in the
variable $x$ or $y$, are interchanged in these equalities, as are the r\^oles
of the variables $x$ and $y$.) 

  What is even less obvious, and actually depends on the
validity of the Heisenberg commutation relation satisfied by $Q$ and
$P$, is that these curves are not only pairwise equal, but in fact they
{\it all} coincide, since the pairs of matrices 
$\ds{(\m{D_1}^N (x)), \ \m{\un D_1}^N(x))}$ and 
$\ds{(\m{D_2}^N (y)), \ \m{\un D_2}^N(y))}$
are conjugate to each other (Theorem \ref{system_duality}), so we also have
the equalities:
\bea
&&\det\le[y\1_{d_2+1 }-{\m{D_1}^N}_ (x) \ri]  =
\det\le[y\1_{d_2+1} - {\m{\un D}^N}_1 (x)\ri]
  \\
&&  \det\le[x\1_{d_1+1 }-{\m{D}^N}_2(y) \ri]  =
 \det\le[x\1_{d_1+1} -  {\m{\un D}^N}_2 (y)\ri]\ .
\eea
Moreover, the transformations relating them are $x$ and $y$ independent, and
are just those  defined by the matrices entering in the generalized
Christoffel--Darboux  relations:
\be
\Amat^N \m{D_1}^N(x) = \m{\un D_1}^N(x)\Amat^N,  \quad
\Bmat^N \m{D_2}^N(y) = \m{\un D_2}^N(y)\Bmat^N \ .
\ee
These same matrices also relate the system of deformation equations,
where they enter as gauge transformations depending on the deformation
parameters $\{u_K, v_J\}$.

   The key to proving all these results lies in noting (Theorem
\ref{maintheorem}) that, as a consequence of the differential equations and
recursion relations satisfied by the $\psi_n$'s, $\phi_n$'s and their
Fourier-Laplace transforms, the following quantities are in fact independent of
all the variables $N$, $x$,  $y$, $\{u_K\}_{K=1\dots d_1+1}$ 
and  $\{v_J\}_{J=1\dots d_2+1}$. 
\be
 \tilde f_N(y) :=
\le(\m{\un{\tilde\Psi}}^{N\!-\!1} (y), \Bmat^N \m{\tilde\Phi}_N(y)\ri)\ , 
\qquad
 \tilde g_N(x) :=
\le(\m{\un{\tilde\Phi}}^{N\!-\!1} (x), \Amat^N \m{\tilde\Psi}_N(x)\ri)  \ ,
\ee
where 
$\ds{\m{\tilde\Psi}_N(x)}$, $\ds{\m{\tilde\Phi}_N(y)}$,
$\ds{\m{\un{\tilde\Psi}}^{N\!-\!1}(y)}$, 
$\ds{\m{\un{\tilde\Phi}}^{N\!-\!1} (x)}$ are {\em any} solutions of the full
system of difference-differential-deformation equations.
This allows us to conclude that there exist compatible sequences of 
fundamental solutions $\ds{\m{\bf \Psi}_{N}(x)}$ and $\ds{\m{\bf \Phi}_{N}(y)}$
of the above difference-differential-deformation equations, and
the corresponding equations for the Fourier-Laplace transformed quantities
$\ds{\m{\bf\un\Phi}^{N-1}}$, $\ds{\m{\bf\un\Psi}^{N-1}}$ such that
\be
\left(\m{\bf\un\Phi}^{N-1} ,  \Amat^N  \m{\bf\Psi}_N)\right)\equiv \1\ ,\qquad
\left(\m{\bf\un\Psi}^{N-1} ,  \Bmat^N  \m{\bf\Phi}_N\right) \equiv \1 \ ,
\ee
for all values of $\{N, x, y, u_J, v_K\}$.

This fact may be viewed as a form of the bilinear identities implying the
existence of $\tau$-functions \cite{UT, AvM2}. The development of this 
relation, and its connection with the isomonodromic deformation equations, 
which requires a study of the formal asymptotics of the fundamental solutions 
near $x=\infty$ and $y=\infty$ will be left to a later work \cite{BEH2}.

In the appendix, all the above results are extended to the sequences of 
multiorthogonal functions that replace the biorthogonal quasi-polynomials
in the multi-dimensional case associated to the multi-matrix model
discussed above.

\medskip

\section{Biorthogonal Polynomials}
\subsection{Biorthogonality measure}

\indent
We first consider  sequences of biorthogonal polynomials with respect to the
measure arising in the study of the two-matrix model discussed in the 
introduction. Consideration of the recursion relations obtained by 
multiplication by or derivation with respect to the independent variable  gives
rise to representations of the Heisenberg relations (string equation) in 
terms of pairs of semi--infinite matrices.  However, most of the results
obtained here may also be shown valid in the fully infinite case.

Let us fix two polynomials, which we refer to  as the ``potentials'',
\bea
V_1(x) =  \sum_{K=1}^{d_1+1}\frac {u_K}{K} x^K\ , \qquad
V_2(y) =  \sum_{J=1}^{d_2+1}\frac {v_J}{J} y^J\ .
\eea 
The coupling constants are normalized in a convenient way so that the
derivatives are 
\be
V_1'(x) = \sum_{K=1}^{d_1+1} u_K x^{K-1}\ .
\ee
We may  define two sequences of mutually orthogonal monic  polynomials
$\pi_n(x),\sigma_n(y)$ of degree $n$ such that 
\bea
&& \int_{\Gamma_x} {\rm d}x \int_{\tilde\Gamma_y} {\rm d}y \,\pi_n(x)\sigma_m(y) {\rm
e}^{-V_1(x)-V_2(y) +xy} = h_n\delta_{mn} \ ,\\
&& \pi_n(x) = x^n + \dots\ , \qquad \sigma_n(y) = y^n+ \dots \ .
\eea 
In order that the integrals be convergent, one should suitably define the two closed 
contours of integration $\Gamma_x, \tilde\Gamma_y$. If we 
require that these be the real axis, the degrees of the potentials must be even 
with the leading coefficient having positive real part.  In applications of 
random matrices to string theory however, the integral is not convergent on 
the real axis, and the contour should approach $\infty$ in some appropriate 
Stokes sector in the complex plane. 

  It is more convenient to deal with the quasi-polynomials defined by 
\be
 \psi_n(x) = \frac 1{\sqrt{h_n}} \pi_n(x){\rm e}^{-V_1(x)} \ , \qquad
\phi_n(y) = \frac 1{\sqrt{h_n}} \sigma_n(y) {\rm e}^{-V_2(y)}\ ,
\label{defpsinphin}
\ee
and their Fourier--Laplace transforms
\be 
\un\psi_n(y) = \int_{\Gamma_x}{\rm d}x\, {\rm e}^{xy} \psi_n(x) \ , \qquad
\un\phi_n(x) = \int_{\Gamma_y}{\rm d}y\, {\rm e}^{xy} \phi_n(y)\ .
\ee 
The choice of normalization is somewhat arbitrary. We have here chosen the most 
``symmetric'' one, in which the leading coefficients in the various recursion
relations are the same.  In this notation the orthogonality relations take on a 
simpler form.  
\be
\int {\rm d}x \,\psi_n(x)\un\phi_m(x) = \int {\rm d} y\,
\un\psi_n(y)\phi_m(y) = \int\int {\rm d}x{\rm d}y\, \psi_n(x)
\phi_m(y){\rm e}^{xy} = \delta_{mn}\ .
\label{ortho}
\ee
(We suppress for the present the specification of the contour of
integration.) We shall think of the spaces spanned by the $\psi_n(x)$'s and
$\phi_n(y)$'s as infinite graded spaces in duality through the pairing in
eq. (\ref{ortho}).  It can easily be seen, using these relations and
integration by parts, that multiplication of the $\psi_n$'s by $x$
produces a linear combination of $\psi_m$'s with $n-d_2\leq m\leq n+1$
and multiplication of the $\phi_n$'s by $y$ produces a linear combination of 
$\phi_j$'s with $n-d_1\leq j\leq n+1$.  Moreover it is clear (through 
integration by parts) that multiplication of the $\psi_n$'s by $x$ is dual to 
application of $\pa_y$  to the $\phi_n$'s and vice-versa.

\subsection{Recursion relations and generalized Christoffel--Darboux formulae}

We denote by $Q$ and $P$ the semi-infinite matrices which implement 
multiplication and differentiation by $x$ on the space spanned by the $\psi_n$ 
quasi-polynomials. Introducing the semi-infinite column vectors 
\be
\ds{\m{\Psi}_\infty :=
[\psi_0,\dots,\psi_n,\dots]^t} \quad  {\rm and}  \quad 
\ds{\m{\Phi}_\infty := [\phi_0,\dots,\phi_n,\dots]^t} \ ,
\ee
the above remarks imply that 
\bea
&& x\,\m{\Psi}_\infty := Q \m{\Psi}_\infty \ ,\qquad
\frac \pa{\pa x}  \m{\Psi}_\infty := P \m{\Psi}_\infty  \ ,\cr  
&& y\, \m{\Phi}_\infty=  -P^t  \m{\Phi}_\infty \ , \qquad
\frac \pa{\pa y}  \m{\Phi}_\infty = -Q^t \m{\Phi}_\infty  \ ,
\label{recursions}
\eea
where $P$ and $Q$ are semi-infinite matrices of the form 
\bea
&& Q :=\le[
\begin{array}{ccccc}
 \a_0(0) &  \gamma(0) & 0 & 0 &\cdots \\
 \a_1(1) & \a_0(1) &\gamma(1) & 0 &\cdots \\
  \vdots &
\ddots
 &\ddots
 &  
\ddots
 &  \ddots
\cr
\a_{d_2}(d_2) & \a_{d_2\!-\!1}(d_2) & \cdots & \a_0(d_2)& \gamma(d_2)
 \cr
0 &
\ddots&
\ddots&
\ddots&
\ddots
\end{array}  
\ri]\label{Qdef}\\
 && -P := \le[
\begin{array}{ccccc}
 \b_0(0) &   \b_1(1) & \cdots & \b_{d_1}(d_1)& \cdots\\
\gamma(0) & \b_0(1) & \b_1(2)&\ddots& \b_{d_1}(d_1\!+\!1)\\
0 & \gamma(1) & \b_0(2) & \ddots &\!\!\!\!\!\!\!\!\!\!\!\!\!\!\!\!\ddots \\
0 &0&  \gamma(2) & \b_0(3) &\!\!\!\!\!\!\!\!\!\!\!\!\!\!\!\!\ddots\\
\vdots& \ddots & \ddots &\ddots &\!\!\!\!\!\!\!\!\!\!\!\!\!\!\!\! \ddots
\end{array}  
\ri]
\label{Pdef}
\eea
satisfying  the string equation.
\be
[P,Q]=\1\ .\label{comm}
\ee
The fact that both matrices have a finite band size as indicated in
(\ref{Qdef}), (\ref{Pdef})  follows from the fact that they are related 
polynomially to each other through the potentials $V_1$ and $V_2$ as follows.
\bl
\label{PQlem}
The two matrices $P$ and $Q$ satisfy the following relations
\bea
&& \le(P + V_1'(Q)\ri)_{\geq 0} =0\ , \label{integrCCRP}\\
&&\le(-Q^t+V_2'(-P^t)\ri)_{\geq 0} =0\ ,\label{integrCCRQ}
\eea
where the subscript $_{\geq 0}$ means the part above the main diagonal
(main diagonal included).
\el
{\bf Proof}. It is obvious that 
\bea
\le[Q\m{\Psi}_\infty\ri]_n &=& x\psi_n(x) = x\frac 1{\sqrt{h_n}} \pi_n(x){\rm e}^{-V_1(x)} \cr
&=& \sqrt{\frac{h_{n+1}}{h_n}} \psi_{n+1}(x) + \hbox
{ lower terms.}\label{one}\\
\le[\le(P+V_1'(Q)\ri)\m{\Psi}_\infty\ri]_n &=& \le(\frac \pa{\pa x} 
+ V_1'(x) \ri) \psi_n(x)  = \frac 1{\sqrt{h_n}}   \pi'_n(x){\rm e}^{-V_1(x)} \cr
&=& n\sqrt{\frac{h_{n-1}}{h_n}} \psi_{n-1}(x) + \hbox { lower terms.}\label{two}
\eea
Eq. (\ref{one}) means that $Q$ has only one diagonal above the main
diagonal with entries given by
\be
\gamma(n) = \sqrt{h_{n+1}/h_n} \ .
\ee
Eq. (\ref{two}) then implies that eq. (\ref{integrCCRP}) holds and $P$ has 
$d_1$ diagonals above the main one.  Repeating the argument for the 
$\phi_n$ quasi-polynomials similarly shows that $-P^t$ (i.e. multiplication by 
$y$) has one diagonal above the main one, eq. (\ref{integrCCRQ}) holds and
$-Q^t$  (i.e. differentiation by $y$) has $d_2$ upper diagonals. 
This proves the lemma and also shows that $P$ and $Q$ are of the finite 
 band sizes indicated in (\ref{Qdef}), (\ref{Pdef}).  \hfill Q.E.D.


Eqs.  (\ref{recursions}) are just equivalent to the following set of
recursion relations  
\bea
&& \le[Q\m{\Psi}_\infty\ri]_n  =x\,\psi_n(x) =  \gamma(n)\psi_{n+1}(x)
 +\sum^{d_2}_{j=0} \a_j(n)\psi_{n-j}(x)  \label{recrel1}\\
&&\le[P\m{\Psi}_\infty\ri]_n = \frac \pa{\pa x} \psi_n(x) = 
 -\gamma(n-1)\psi_{n-1}(x)  - \sum^{d_1}_{j=0} \b_{j}(n+j)\psi_{n+j}(x)
\label{recrel2} \\
&&\le[-P^t \m{\Phi}_\infty\ri]_n = y\, \phi_n(y)=
\gamma(n)\phi_{n+1}(y)  +\sum^{d_1}_{j=0}
\b_j(n)\phi_{n-j}(y) \label{recrel3} \\
&&\le[-Q^t \m{\Phi}_\infty\ri]_n = \frac \pa{\pa y} \phi_n(y) =
-\gamma(n-1)\phi_{n-1}(y)   - \sum^{d_2}_{j=0}
\a_{j}(n+j)\phi_{n+j}(y) \ ,\label{recrel4}
\eea
 In the following we also define
\bea
&& \a_{-1}(n):=\gamma(n)=:\beta_{-1}(n)\ , \quad
\a_{j}(n):= 0 ,\ \forall j\not \in [-1,d_2] \ , \quad
 \b_{j}(n):= 0 \ ,\ \forall j\not \in [-1,d_1].
\eea

Defining similarly the semi-infinite row vectors consisting of the 
Fourier-Laplace transformed functions 
\be
{\ds{\m{\un\Psi}_\infty } := [\un\psi_0, \cdots,\un\psi_n, \cdots]} \quad
{\rm and} \quad {\ds{\m{\un\Phi}_\infty } := [\un\phi_0, \cdots,\un\phi_n, \cdots]}
\ ,
\ee
it follows from the dual pairing (\ref{ortho}),  and integration by parts 
that:
\bl
\bea
\le[- \m{\un\Psi}_\infty P^t\ri]_n &=& y\un\psi_n(y) =
\sum_{j=-1}^{d_1} \b_{j}(n+j)\un\psi_{n+j}(y)\ ,
\label{recrelbar1} \\
\le[\m{\un\Psi}_\infty Q^t\ri]_n &=& \pa_y \un\psi_n(y) =
\sum_{l=-1}^{d_2} \a_{l}(n)\un\psi_{n-l}(y)\ ,  \label{recrelbar2} \\
\le[\m{\un\Phi}_\infty Q\ri]_n &=&x\un\phi_n(x) = 
\sum_{l=-1}^{d_2} \alpha_{l}(n+l)\un\phi_{n+l}(x)\ , \label{recrelbar3}  \\
\le[-\m{\un\Phi}_\infty P\ri]_n &=& \pa_x \un\phi_n(x) = 
\sum_{j=-1}^{d_1} \b_{j}(n)\un\phi_{n-j}(x)  \ .  \label{recrelbar4} 
\eea
\el
\smallskip  
  Now, we introduce the shift matrices
\be
\Lambda:= \le[\matrix { 0&1&0&0&0& \cdots\cr 0&0&1&0&0&\cdots \cr
0&0&0&1&0&\cdots \cr 0&0&0&0&1&\cdots \cr
0&0&0&0&0&\ddots} \ri] \ ,
\qquad
\Lambda^{-1}:=\Lambda^{t}=
\le[\matrix { 0&0&0&0&0& \cdots\cr 1&0&0&0&0&\cdots \cr
0&1&0&0&0&\cdots \cr 0&0&1&0&0&\cdots \cr
0&0&0&0&\ddots&\cdots} \ri]
\ee
(The notation $\Lambda^{-1}$ is a convenient shorthand for the
transpose $\Lambda^t$, but only signifies that $\Lambda^{-1}$ is the
{\em right} inverse of $\Lambda$. It  leads to the abbreviated form
$\Lambda^{-j}:= (\Lambda^t)^{j}$.)  Introducing  the diagonal semi-infinite
matrices 
\bea
\a_j := {\rm diag}(\alpha_j(0),\a_j(1),\dots)\ ,
\qquad \b_j := {\rm diag}(\b_j(0),\b_j(1),\dots)\ ,
\eea
(where we set $\alpha_j(n):=0$ and $\beta_j(n):=0$ when $n<j$), eqs.
(\ref{Qdef}) , (\ref{Pdef}) can be concisely written as 
\bea
 Q := \sum_{j=-1}^{d_2} \Lambda^{-j} \a_{j}\ ;\qquad 
 -P :=  \sum_{j=-1}^{d_1} \Lambda^{j} \b_{j}\ .
\eea

The commutation relation in eq. (\ref{comm})  gives in particular the following
quadratic  relations between the coefficients $\{\a_j,$ $\b_k\}$,
\bea
 \sum_{j=l}^{d_1} \beta_{j}(n+j) \alpha_{j-l-1}(n+j) &=&
\sum_{j=l}^{d_1} \alpha_{j-l-1}(n) \beta_{j}(n+l+1)\ ,\ \  \forall n \,
, \, \forall\,  l \in [0,d_1] \label{heisenbergone}\\ 
\sum_{k=l}^{d_2}
\beta_{k-l-1}(n+k) \alpha_{k}(n+k) &=& \sum_{k=l}^{d_2}
\alpha_{k}(n+l+1) \beta_{k-l-1}(n)\ ,\ \ \forall n \, , \, \forall\, l \in [0,d_2]  \ .
\label{heisenbergtwo}
\eea

Our first objective is to define a set of closed differential and
difference  systems for vectors consisting of finite sequences of 
the functions $\{\psi_n\}$, $\{\phi_n\}$ and their Fourier-Laplace transforms,
equivalent to the systems (\ref{recrel1})-(\ref{recrel4}), and to 
study their properties and relations.  
Consider first  the sequence of functions $\{\psi_n(x)\}$. For these we have a 
multiplicative recursion relation defined by the coefficients $\{\alpha_j(n)\}$
 and a differential recursion relation  defined by the coefficients 
$\{\beta_j(n)\}$. We shall show presently that we can define {\em closed} 
systems of first order linear ODE's for any consecutive sequence of $d_2+1$ 
functions $(\psi_{N-d_2},\dots,\psi_{N})$ (or 
$(\un\phi_{N-1},\dots,\un\phi_{N+d_2-1})$)  with  coefficients that are 
polynomials in $x$.  Similar systems can be constructed for any
sequence $(\phi_{N-d_1},\dots,\phi_N)$ (or
$(\un\psi_{N-1},\dots,\un\psi_{N+d_1-1})$).   We introduce the following
definitions and  notations
\bd
\label{dualwindows}
A {\bf window of size $\mathbf d_1$ or $\mathbf d_2$} is any  subset of 
$d_1$ or $d_2$  consecutive elements of type $\psi_n$, $\un \phi_n$, 
 $\phi_n$ or $\un\psi_n$, with the notations 
\bea
&&\m{\Psi}_N :=[\psi_{N\!-\!d_2},\dots,\psi_{N}]^t \ , \quad  N\geq d_2, \qquad
\m{\Phi}_N :=[\phi_{N\!-\!d_1},\dots,\phi_{N}]^t  \ ,\quad  N\geq d_1 \\
&&\m{\Psi}^N :=[\psi_{N},\dots,\psi_{N\!+\!d_1}]^t \ , \quad  N\geq 0,   \qquad
\m{\Phi}^N :=[\phi_{N},\dots,\phi_{N\!+\!d_2}]^t \ , \qquad N\geq 0 \\ 
&&\m{\un\Psi}_N :=[\un\psi_{N\!-\!d_2},\dots,\un\psi_{N}] \ , \quad  N\geq d_2, 
\qquad
\m{\un\Phi}_N :=[\un\phi_{N\!-\!d_1},\dots,\un\phi_{N}]  \ ,\quad  N\geq d_1 \\
&&\m{\un\Psi}^N :=[\un\psi_{N},\dots,\un\psi_{N\!+\!d_1}]\ , \quad  N\geq 0,   
\qquad 
\m{\un\Phi}^N :=[\un\phi_{N},\dots,\un\phi_{N\!+\!d_2}]\ , \qquad N\geq 0 \ .
\eea
\ed
Notice the difference in the positioning of the windows for the vectors
constructed from the $\psi_n$'s and the $\phi_n$'s, and the fact that the
barred quantities are defined to be row vectors while the unbarred ones
are column vectors.
\bd
For any $N$ for which these are defined, the pairs of windows
 ($\ds{\m{\Psi}_N}, \  \ds{\m{\un{\Phi}}^{N-1} }$)
 as well as   ($\ds{\m{\Phi}_N}, \ \ds{\m{\un{\Psi}}^{N-1}}$) of dimensions
$d_2+1$ and $d_1+1$, respectively, will be called {\bf dual windows}.
\ed
The reason for identifying these particular windows as dual will appear in the
sequel.
\par
Let us now consider the kernels
\bea
&&{\m{K}^N}_{11}(x,x') := \sum_{n=0}^{N-1}  \psi_n(x) \underline\phi_n(x')\ ,
\qquad
{\m{K}^N}_{22} (y',y) := \sum_{n=0}^{N-1} \underline \psi_n(y')
\phi_n(y)\ , \\
&&{\m{K}^N}_{12}(x,y) := \sum_{n=0}^{N-1}  \psi_n(x) \phi_n(y)\ ,
\qquad
{\m{K}^N}_{21} (y',x') := \sum_{n=0}^{N-1} \underline \psi_n(y')
\underline \phi_n(x')\ , 
\eea
that appear in the computation of correlation functions for
$2$--matrix models \cite{eynardmehta}.

 Define  the following pair of matrices, which will play an important
r\^ole in what follows:
\bea
&& \Amat^N := \le[
\begin{array}{cccc|c}
0&0&0&0&\!\!-\!\gamma(\!N\!\!-\!1\!)\!\!\cr\hline
\a_{d_2}\!(\!N\!)& \cdots & \a_{2}(\!N\!)& \a_1(\!N\!)& 0\cr
0& \a_{d_2}\!(\!N\!\!+\!1\!) & \cdots & \a_1(\!N\!+\!1\!)& 0\cr
0&0&\a_{d_2}\!(\!N\!\!+\!2\!) &\cdots & 0\cr
0&0&0&\a_{d_2}\!(\!N\!\!+\!d_2\!-\!1\!)&0
\end{array}
\ri]\ ;\\[10pt]
&& \Bmat^N := \le[
\begin{array}{cccc|c}
0&0&0&0&\!\!-\!\gamma(\!N\!\!-\!1\!)\!\!\cr\hline
\b_{d_1}\!(\!N\!)& \cdots & \b_{2}(\!N\!)& \b_1(\!N\!)& 0\cr
0& \b_{d_1}\!(\!N\!\!+\!1\!) & \cdots & \b_1(\!N\!+\!1\!)& 0\cr
0&0&\b_{d_1}\!(\!N\!\!+\!2\!) &\cdots & 0\cr
0&0&0&\b_{d_1}\!(\!N\!\!+\!d_1\!-\!1\!)&0
\end{array}
\ri]
\eea

For any $N$, the recursion relations (\ref{recrel1}), (\ref{recrel3}), 
(\ref{recrelbar1}), (\ref{recrelbar3}) and the differential relations
(\ref{recrel2}), (\ref{recrel4}), (\ref{recrelbar2}), (\ref{recrelbar4})
imply that the following generalized Christoffel--Darboux formulae, as
well as their ``differential'' analogs are satisfied.
\bp
\label{generalDC}
Generalized Christoffel--Darboux relations:
\bea
(x-x') {\m{K}^N}_{11}(x,x') & =&  \gamma(N-1) \psi_N\un\phi_{N-1} -
\sum_{j=1}^{d_2} \sum_{k=0}^{j-1} \a_j(N+k) \un\phi_{N+k} \psi_{N+k-j}  \cr
&=&
-  \le(\m{\un\Phi}^{N\!-\!1} (x'),\Amat^N  \m{\Psi}_{N} (x)\ri) \ ,
\label{DarbChris1}\\
  (y'-y){\m{K}^N}_{22}(y',y)& =& -\gamma(N-1)\un\psi_{N-1} \phi_{N} + \sum_{j=1}
^{d_1} \sum_{k=0}^{j-1} \beta_{j}(N+k) \un \psi_{N+k} \phi_{N-j+k}  \cr
&=&
 \le(\m{\un\Psi}^{N\!-\!1} (y'), \Bmat^N  \m{\Phi}_{N} (y)\ri) \ . 
\label{DarbChris2}
\eea
``Differential'' generalized Christoffel--Darboux relations:
\bea
 \le(\pa_{x'}+\pa_x\ri){\m{K}^N}_{11}(x',x) & = &
- \le(\m{\un\Phi}_{N} (x'),(\Bmat^N)^t \m{\Psi}^{N\!-\!1} (x)\ri) \ , 
\label{DarbChrisDiff1}  \\ 
  \le(\pa_{y'}+\pa_y\ri){\m{K}^N}_{22}(y',y) & = & 
-\le(\m{\un\Psi}_N (y'), (\Amat^N)^t \m{\Phi}^{N\!-\!1} (y)\ri)  \ .
\label{DarbChrisDiff2} 
\eea
\ep
{\bf Proof}. Use the relations (\ref{recrel1})-(\ref{recrel4}),
(\ref{recrelbar1})-(\ref{recrelbar4})  and simplify the telescopic sums by 
cancellation of common terms. \hfill Q.E.D.

   Although it will not be needed in the remainder of this paper, for the sake 
of completeness, we also include the following analogous result for the 
kernels $K_{12}$ and $K_{21}$, which may  be similarly derived. It is 
related to the above by applying Fourier-Laplace transforms with respect
to one of the variables.

\bp
\bea
(x+\pa_y) {\m{K}^N}_{12}(x,y) &=&
 - \le({\m{\Phi^t}^{N\!-\!1}} (y), \Amat^N \m{\Psi}_{N} (x)\ri) \ , 
\label{DarbChris121}\\
  (y + \pa_x){\m{K}^N}_{12}(x,y) &=&
-\le({\m{\Psi^t}^{N\!-\!1}}(y'), \Bmat^N \m{\Phi}_{N} (y)\ri) \ , 
\label{DarbChris122} \\ 
\le(y'-\pa_{x'}\ri){\m{K}^N}_{21}(y',x') &=&
 \le(\m{\un\Psi}^{N\!-\!1} (y'),\Bmat^N \m{\un\Phi}_{N} (x')\ri) \ , 
\label{DarbChrisDiff211}  \\
  \le(\pa_{y'}-x'\ri){\m{K}^N}_{21}(y',x') &=&
- \le(\m{\un\Phi}^{N\!-\!1} (x'), \Amat^N {\m{\un\Psi}_{N}}^t (y')\ri) \ . 
\label{DarbChrisDiff212} 
\eea
\ep

\subsection{Folding}

We now introduce the sequence of companion--like  matrices $\ds{\A_N(x)}$
and $\ds{ \B_N(y)}$ of sizes $d_2+1$ and $d_1+1$, respectively.
\bea
\A_N(x) & := 
\le[\begin{array}{cccc} 0  & 1 & 0 &\!\!\! \!\!\!\!\!\! \!\!\! 0 \cr
0 & 0 & \ddots &\!\!\! \!\!\!\!\!\! \!\!\!0\cr
0 & 0 & 0 &\!\!\! \!\!\!\!\!\! \!\!\!1\cr
\!\!\frac {-\alpha_{d_2}(N)} {\gamma(N)}\!\! &
\cdots
&\!\! \frac {-\alpha_1(N)}
{\gamma(N)} \!\! &\!\!  \frac{(x-\alpha_0(N))}{\gamma(N)} \!\!  
\end{array}\ri] \ ,  \quad N\geq d_2 \ ,  \label{aNdef}\\
{\B_{N}}(y) & := \le[\begin{array}{cccc} 
  0 & 1& 0 &\!\!\! \!\!\!\!\!\! \!\!\!0 \cr
  0 & 0 & \ddots  &\!\!\! \!\!\!\!\!\! \!\!\!0 \cr
  0 & 0 & 0 &\!\!\! \!\!\!\!\!\! \!\!\! 1 \cr
\!\! \frac {-\b_{d_1}(N)} {\gamma(N)} \!\! &
\cdots
 &\!\! 
\frac { -\b_1(N)} {\gamma(N)} \!\!     &\!\!
\frac{(y-\b_0(N))}{\gamma(N)} 
\end{array}\ri] \ ,  \quad N\geq d_1  \ .   \label{bNdef}
\eea
We then have the following: 
\bl \label{abNrecursions} 
The sequence of matrices  $\ds{\m{\bf a}_N}$, $\ds{\m{\bf b}_N}$ implement the 
shift  $N\mapsto N+1$ in the windows of quasi-polynomials in the sense that  
\be \A_N\m{\Psi}_N(x) =
\m{\Psi}_{N\!+\!1}(x)\ , \qquad 
\B_N\m{\Phi}_N(y) =
\m{\Phi}_{N\!+\!1}(y) \label{recs1} \ ,
\ee
 and in general 
\be
\m{\Psi}_{N\!+\!j} = \A_{N+j-1}\cdots\A_{N}
\m{\Psi}_N\ ,\qquad \m{\Phi}_{N\!+\!j} = \B_{N+j-1}\cdots\B_{N}
\m{\Phi}_N\ .
\label{folding}
\ee
 \el
 {\bf Proof}.  This is nothing but a matricial form of the
sequence of recursion relations (\ref{recrel1}), (\ref{recrel3}) 
expressing the higher order polynomials as linear combinations of a fixed
subset with polynomial coefficients. \hfill Q.E.D.

We will refer to this process of expressing any $\psi_n(x)$ by means of linear 
combinations of elements in a specific window with polynomial coefficients as 
{\em folding} onto the specified window. 

The determinants of the matrices $\ds{\A_N}$ and $\ds{\B_N}$ are easily 
computed to be 
\be
 \det(\A_N) = (-1)^{d_2+1} \a_{d_2}(N)/\gamma(N)\ ,\qquad \det(\B_N) =
(-1)^{d_1+1} \b_{d_1}(N)/\gamma(N) \ .
\ee
From  eqs. (\ref{integrCCRP}), (\ref{integrCCRQ}) we find the relations
\bea
\a_{d_2}(N) = v_{d_2+1} \prod_{j=1}^{d_2} \gamma(N-j)\ ;\qquad 
\b_{d_1}(N) = u_{d_1+1} \prod_{j=1}^{d_1} \gamma(N-j)\ .
\label{integratedCCR}
\eea
Since the coefficients $\gamma(N)$ are the square roots of the ratios
of normalization factors, they cannot vanish for any $N$, and neither can
 $\a_{d_2}(N)$ or $\b_{d_1}(N)$, since the deformation parameters 
$u_{d_1+1}$,  $v_{d_2+1}$ are the leading coefficients of the polynomials
$V_1(x)$, $V_2(y)$ and hence also may not vanish.  It follows that the 
matrices $\ds{\A_N}$ and $\ds{\B_N}$ are all invertible. We denote their 
inverses as follows
\bea
  \A^{N} &:=& \le[\A_{N}\ri]^{-1} = \left [\begin {array}{cccc}
{\frac {-\alpha_{{d_2\!\!-\!1}}\!(N)}{\alpha_{{d_2}}(N)} }&\cdots &
{\frac {x-\alpha_{{0}} (N)}{\alpha_{{d_2}}(N)}}&
{\frac {-\gamma(N)}{\alpha_{{d_2}}(N)}}
\\\noalign{\medskip}1&0&0&0\\\noalign{\medskip}0&\ddots&0&0
\\\noalign{\medskip}0&0&1&0\end {array}\right ]
 \   \\
\B^{N} &:=& \le[\B_{N}\ri]^{-1}= \left [\begin {array}{cccc}
{\frac {-\b_{{d_1\!\!-\!1}}\!(N)}{\b_{{d_1}}(N)} }&\cdots &{\frac {y-\b_{{0}}
(n)}{\b_{{d_1}}(N)}}&{\frac {-\gamma(N)}{\b_{{d_1}}(N)}}
\\\noalign{\medskip}1&0&0&0\\\noalign{\medskip}0&\ddots&0&0
\\\noalign{\medskip}0&0&1&0\end {array}\right ]
\eea
The shifts $N\to N-1$ are thus implemented by the inverse matrices $\ds{\A^N}$,
 $\ds{\B^N}$, and the folding may take place in either direction with 
respect to polynomial degrees.

\subsection{Folded linear differential systems}

We now define the following sequences of finite diagonal matrices  
\bea
 &&
\stackrel{N}\a_j := 
{\rm diag}\le[\a_j(N+j-d_1),\a_j(N+j-d_1+1),\dots,\a_j(N+j) \ri] \ ,\
j=-1, \dots d_2\ , \label{alphajN}\\
&&
\stackrel{N}\b_j :=
{\rm diag}\le[\b_j(N+j-d_2),\b_j(N+j-d_2+1),\dots,\b_j(N+j) \ri]\ ,\ 
j=-1,\dots d_1
\ .\label{betajN}
\eea
Recall that $\a_{-1}(n)= \gamma(n)=\b_{-1}(n)$ by our previous conventions, 
but the diagonal matrices $\ds{{\m\a^N}_{-1}}$ and $\ds{{\m\b^N}_{-1}}$
differ in  dimensions. When  the context  leaves no doubt as to the dimension
we will  write them as  
\be 
 \stackrel{N}\a_{-1}  =\stackrel{N}{\gamma} :=
{\rm diag}\le[\gamma(N-d_1-1), \dots, \gamma(N-1) \ri] 
\ee
 or
 \be
\stackrel{N}\b_{-1} = \stackrel{N}{\gamma} := {\rm
diag}\le[\gamma(N-d_2-1),\gamma(N-d_2),\dots, \gamma(N-1) \ri] \ .
\ee
In either case, we denote the inverse matrix as
\be
\m{\gamma}_{N} := (\stackrel{N}{\gamma})^{-1} \ .
\ee
We can now  give  the closed differential systems referred to previously.
\bl
\label{D12}
The windows of quasi-polynomials $\ds{\m{\Psi}_N}$,
$\ds{\m{\Phi}_N}$ satisfy the following differential systems
\bea
\frac \pa{\pa x} \m{\Psi}_N &=& -\m{D_1}^N (x)
 \m{\Psi}_N  \ ,  \quad N\ge d_2+1 \ , \label{xDE_PsiN}\\
\frac \pa{\pa y} \m{\Phi}_N &=& -\m{D_2}^N (y)
 \m{\Phi}_N \ , \quad N\ge d_1+1 \ ,\label{yDE_PhiN}
\eea
where 
\bea
 &&
 \m{D_1}^N(x) := 
 \stackrel{N}{\gamma}\m{\A}^{N-1} +
\stackrel{N}{\b_0} + \sum_{j=1}^{d_1}\stackrel{N}{\b_j}\A_{N+j-1}
\A_{N+j-2}\cdots
\A_N\ \in\ gl_{d_2+1}[x]  \ . \label{defD1}\\
 && \m{ D_2}^N(y) :=
 \stackrel{N}{\gamma} \m{\B}^{N-1} +
\stackrel{N}{\a_0} + \sum_{j=1}^{d_2}\stackrel{N}{\a_j}  {\B_{N+j-1}}
{\B_{N+j-2}}\cdots{\B_N} \ \in\ gl_{d_1+1}[y]\ .  \label{defD2}
\eea 
These, taken together for all $N$ are equivalent to the relations
(\ref{recrel2}), (\ref{recrel4}) when the recursion relations (\ref{recs1}),
(\ref{folding}) are taken into account.
\el
{\bf Proof}. 
Consider the case of the $\psi_n$'s.  The differential relations (\ref{recrel2}) 
may be written  by stacking them in a window of size $d_2+1$, as follows
\be
\frac \pa{\pa x} \m{\Psi}_N = -\m{\gamma}^N \m{\Psi}_{N\!-\!1}
- \sum_{j=0}^{d_2} \m{\b_j}^N \m{\Psi}_{N\!+\!j} \ .
\ee
Using the folding relations eq. (\ref{folding}), we immediately  obtain
(\ref{xDE_PsiN}), (\ref{defD1}).  The same procedure applied to eq. (\ref{recrel4})
yields (\ref{yDE_PhiN}), (\ref{defD2}). \hfill Q.E.D.

We can repeat a similar procedure for the sequences 
 $\{\un\psi_n(y)\}_{n\in \N}$ and $\{\un\phi_n(x)\}_{n\in \N}$. 
The corresponding windows are represented as row vectors ${\ds\m{\un\Psi}^{N}}$ 
and ${\ds\m{\un\Phi}^{N}}$ since their components are naturally dual to 
the $\phi_n$'s and $\psi_n$'s respectively. The matrices defining the 
relevant folding are now
\bea
 \m{\un\A}^{N} & := \le[
\begin{array}{cccc} 
  \frac{x\!-\!\a_0\!(N)}{\gamma(N\!-\!1)}&1&0&0\\[5pt]
 \frac{-\a_1\!(N\!+\!1)}{\gamma(N\!-\!1)}&0&
^{\ds{^{\ds{\cdot}}}}\hbox{}
 ^{\ds{\cdot}} \cdot&0\\
 \vdots &0&0&1\\[5pt]
   \frac{-\a_{d_2}\!(N\!+\!d_2)}{\gamma(N\!-\!1)}&0&0&0
\end{array} \ri]\ , 
\label{abarN} \\
 \m{\underline\B}^{N} & := \le[
\begin{array}{cccc}
 \frac{y\!-\!\b_0\!(N)}{\gamma(N\!-\!1)}&1&0&0\\[5pt]
 \frac{-\b_1\!(N\!+\!1)}{\gamma(N\!-\!1)}&0&
 ^{\ds{^{\ds{\cdot}}}}\hbox{}
 ^{\ds{\cdot}} \cdot&0\\
 \vdots &0&0& 1\\[5pt]
  \frac{-\b_{d_1}\!(N\!+\!d_1)}{\gamma(N\!-\!1)}&0&0&0
\end{array} \ri] \ ,  \label{bbarN}
\eea
and we again denote their inverses as
\be
\m{\un\A}_N:=\le[\m{\un\A}^N \ri]^{-1} \ ,  \qquad
\m{\un\B}_N:=\le[\m{\un\B}^N \ri]^{-1} \ .
\ee
As previously , we now have:
\bl
The sequence of matrices ${\ds \{\m{\un\A}^N\}}$ ${\ds \{\m{\un\B}^N\}}$ 
implement the shift $N\mapsto N-1$
\be
\m{\un\Psi}^{N\!-\!1}  =
\m{\un\Psi}^{N}\m{\un\B}^N  \ ;\qquad 
\m{\un\Phi}^{N\!-\!1}  =
\m{\un\Phi}^{N}\m{\un\A}^N \ . \label{recs_bar1}
\ee
\el
Similarly to the diagonal matrices  $\ds{\m{\a_j}^N,\ \m{\b_j}^N}$, we 
define the matrices $\ds{\m{\un\a_j}^N}$ and $\ds{\m{\un\b_j}^N}$ as
\bea
&& \m{\underline \alpha_j}^{N} :={\rm
diag}\le(\a_j(N),\a_j(N+1),\dots,\a_j(N+d_1)\ri)\ , \ \ j=-1,\dots,d_2
\label{alphabarjN} \\
&& \m{\underline \b_j}^{N} :={\rm
diag}\le(\b_j(N),\b_j(N+1),\dots \b_j(N+d_2)\ri)\ ,\ \ j=-1,\dots,d_1\ .
\label{betabarjN}
\eea
As before we have the two definitions 
\bea
&& {\m{\underline \a}^{N}}_{-1}:=\m{\underline \gamma}^{N} :={\rm
diag}\le(\gamma(N),\gamma(N+1),,\dots\gamma(N+d_1)\ri)\\
&& {\m{\underline \b}^{N}}_{-1}:=\m{\underline \gamma}^{N} :={\rm
diag}\le(\gamma(N),\gamma(N+1),\dots,\gamma(N+d_2)\ri)\ ,
\eea
which will be used if there is no ambiguity regarding dimensions.

By repeating a procedure similar to what led to the differential systems in 
Lemma \ref{D12}, we find: 
\bl
\label{_D12}
The dual windows of Laplace--transformed quasi-polynomials 
$\ds{\m{\un\Psi}^{N-1}} $, $\ds{\m{\un\Phi}^{N-1}}$ satisfy the following 
differential systems
\bea
\frac \pa{\pa y}\m{\un\Psi}^{N\!-\!1}(y) & =
\ds{\m{\un\Psi}^{N\!-\!1}(y) {\m{\un D}^N}_2(y)}\ ,
\quad N\ge d_1+1 \ , \label{yDE_PsibarN} \\
\frac \pa{\pa x}\m{\un\Phi}^{N\!-\!1}(x) & =
\ds{\m{\un\Phi}^{N\!-\!1}(x) {\m{\un D}^{N}}_1(x)} \ , 
\quad N\ge d_2+1 \ , \label{xDE_PhibarN}
\eea
where
\bea
{\m{\un D}^N}_2(y) &:= \ds{\m{\un\B}_{N}
\m{\underline \gamma}^{N-1}  + {\m{\underline\a_0}^{N-1}}
+\sum_{j=1}^{d_2}
\m{\un\B}^{N-1}\m{\un\B}^{N-2}\cdots
\m{\un\B}^{N-j}
{\m{\un\a_j}^{N-1}} }  \ , \\
\m{\underline D_1}^N(x) & := \ds{\m{\un\A}_{N}
\m{\underline \gamma}^{N-1}  + {\m{\underline\b_0}^{N-1}}
+\sum_{j=1}^{d_1}
\m{\un\A}^{N-1}\m{\un\A}^{N-2}\cdots
\m{\un\A}^{N-j}
{\m{\underline\b_j}^{N-1}} } \ .
\eea
\el
Summarizing, we have thus obtained four differential systems
\bea
\begin{array}{l|l}
\hbox {Size } (d_2+1)\times (d_2+1) & \hbox {Size } (d_1+1)\times
(d_1+1) \\
\hline
\vrul 
\ds{
\frac \pa{\pa x}\m{\Psi}_{N}  (x) = - \m
{ D_1}^N(x)\m{\Psi}_{N}
(x) }& \ds{
\frac \pa{ \pa y}\m{\underline \Psi}^{N\!-\!1}(y) =
\m{\un\Psi}^{N\!-\!1}(y)\m{\underline
D_2}^N (y)}\\[10pt] 
\ds{\frac \pa{\pa x}\m{\underline \Phi}^{N\!-\!1}
(x) =
\m{\un\Phi}^{N\!-\!1}(x)\m{\underline
D_1}^N (x)} &
\ds{\frac \pa{\pa y}\m{\Phi}_{N}  (y) = - \m
{ D_2}^N(y)\m{\Phi}_{N}(y)}
\end{array} \label{foursystem}
\eea

It should be noted that the two matrices $D_1$ and $\un D_1$ (as
well as $D_2$ and $\un D_2$) have so far only superficial similarities. 
In particular they {\em do not} depend on the same  subsets of the
coefficients $\{\a_j(n)\}$ and  $\{\b_j(n)\}$. On the other hand the pairs 
$(D_1, \un D_2)$ and $(D_2,\un D_1)$ {\em do} depend on the same 
$\alpha_j(n)$'s and $\beta_j(n)$'s although they are of different dimensions. 

\subsection{Deformation equations}
The following Lemma gives the effect of an  infinitesimal  deformation
in the coefficients $\{u_K, v_J\}$  expressed as differential equations for the
(semi)-infinite vectors $\m{\Psi}_\infty$, $\m{\Phi}_\infty$ of biorthogonal
quasi-polynomials (as well as their  Fourier--Laplace transforms) and for the
matrices $P,Q$. (Derivations with different conventions can be found in 
\cite{ZJDFG, AvM1, McLaughlin}.)
\bl
\label{uvPSiPhidefs}
\bea
\pa_{u_K} \m{\Psi}_\infty &=& U^K \m{\Psi}_\infty\ , \label{defUK}\\
\pa_{v_J} \m{\Psi}_\infty &=& - \le(V^J\ri)^t \m{\Psi}_\infty \
,\label{defVJt}\\
 \pa_{u_K} \m{\Phi}_\infty &=& -\le(U^K\ri)^t \m{\Phi}_\infty
\ ,\label{defUKt}\\ 
\pa_{v_J} \m{\Phi}_\infty &=& V^J \m{\Phi}_\infty\ ,
\label{defVJ} \eea
where 
\be
 U^{K} :=-\frac 1 K \le\{ \le[Q^K\ri]_{>0} + \frac 1 2
\le[Q^K\ri]_0\ri\}\ ,\ \  
V^{J} :=-\frac 1 J \le\{ \le[(-P^t)^J\ri]_{>0} + \frac 1 2 
\le[(-P^t)^J\ri]_0\ri\}\ .
\label{UKVJexpr}
\ee
Componentwise these read, 
\bea
&&  \ds{\pa_{u_K} \psi_n(x)= \sum_{j=0}^K U^{K}_j(n) \psi_{n+j}(x)} \ ,
\label{defpsinuK} \\
&& \ds{ 
 \pa_{v_J} \psi_n(x)= - \sum_{j=0}^J V^{J}_j(n-j) \psi_{n-j}(x)} \ ,
\label{defpsinvJ}  \\
&&  \ds{ \pa_{u_K} \phi_n(y)= - \sum_{j=0}^K U^{K}_j(n-j)
 \phi_{n-j}(y)} \ , \label{defphinuK}  \\
&&  \ds{ \pa_{v_J} \phi_n(y)= \sum_{j=0}^J
V^{J}_j(n)\phi_{n+j}(y)}  \label{defphinvJ}  \ ,
\eea
where we have used the notation 
\be
U^K_j(n):= U^K_{n,j+n}\ ,\ \ \ V^J_j(n):=
V^J_{n,j+n}\ .
\ee 
(The same relations hold for the Fourier-Laplace transforms with respect
to the variables $x$ and $y$, since the coefficients do not depend on these
variables.)

Moreover, we have the following equations for the matrices $P,Q$:
\bea
&&\pa_{u_K}Q =  -[Q,U^K] ,\label{deformQu} \\
&& \pa_{v_J} Q =[Q,{V^J}^t]\ ,\label{deformQv} \\
&&  \pa_{u_K} P = -[P,U^K]\ ,\label{deformPu} \\
&& \pa_{v_J} P =  [P,{V^J}^t]\ .\label{deformPv} 
\eea
\el

{\noindent \bf Proof:}
Equations (\ref{defUK}) and (\ref{defVJ}) are just definitions, (\ref{defUKt})
and (\ref{defVJt}) follow from (\ref{ortho}).
Eq. (\ref{UKVJexpr}) is proved in a way similar to Lemma \ref{PQlem}. From
the definitions (\ref{defpsinphin}), one has:  
\be
 \le[U^K \m{\Psi}_\infty \ri]_n = \pa_{u_K} \psi_n(x) = -{1\over 2} {\pa_{u_K}
h_n \over h_n} \psi_n(x) + {1\over \sqrt{h_n}} \pa_{u_K} \pi_n(x) {\rm
e}^{-V_1(x)} - {1\over K} x^K \psi_n(x)  
\label{UKpsiinfty} \ ,
\ee
and
\be
 \le[-\le( U^K \ri)^t \m{\Phi}_\infty \ri]_n = \pa_{u_K} \phi_n(y) = -{1\over
2} {\pa_{u_K} h_n \over h_n} \phi_n(y) + {1\over \sqrt{h_n}} \pa_{u_K}
\sigma_n(y) {\rm e}^{-V_2(y)}  \ . 
\label{UKphiinfty}
\ee
Since $\pa_{u_K} \pi_n(x)$ has a degree lower than $n$, one sees that eq.
(\ref{UKpsiinfty}) implies that 
\be
\le( U^K\ri)_{>0} = -{1\over K} \le( Q^K \ri)_{>0} \ ,
\ee
and similarly eq. (\ref{UKphiinfty}) implies that 
\be
\le( U^K\ri)_{<0} = 0 \ .
\ee
They also imply that the diagonal part must be:
\be
U^K_{n,n} = -{1\over 2}{\pa_{u_K} h_n \over h_n} -{1\over K}  \le( Q^K
\ri)_{n,n} = {1\over 2}{\pa_{u_K} h_n \over h_n} = -{1\over 2K}  \le( Q^K
\ri)_{n,n}  \ , 
\ee 
which proves (\ref{UKVJexpr}).

The equations (\ref{deformQu})--(\ref{deformPv}) follow from multiplying
(\ref{defUK}) and (\ref{defVJt}) by $x$ and (\ref{defUKt}) and (\ref{defVJ}) 
by $y$ and the linear independence of the component functions forming
the vectors   The coefficients of
the expansion must vanish, and that is precisely the relations 
(\ref{deformQu}) to (\ref{deformPv}).
\hfill   Q.E.D.

\smallskip
We also require the ``folded'' version of the deformation equations.
This leads to eight equations giving the action of $\pa_{u_K}$ and $\pa_{v_J}$ 
on $\ds{\m{\Psi}_{N}}$,  $\ds{\m{\un\Psi}^{\!N-\!1}}$, $\ds{\m{\Phi}_{N}}$ and
$\ds{\m{\un\Phi}^{\!N-\!1}}$.
They are introduced in the following lemma, for which we need to define
 diagonal matrices which play  r\^oles  similar to that of the matrices 
defined in (\ref{alphajN}),  (\ref{betajN}) (\ref{alphabarjN}) , 
(\ref{betabarjN}) for the differential equations with respect to $x$ or $y$.
\bea
&&{\m{U}^{N,d}}_{j,K} :=
 {\rm diag}(U^{K}_j(N\!-\!d),\dots,U^{K}_j(N)) \ ,\\ 
&&{\m{V}^{N,d}}_{j,J}
:={\rm diag}(V^{J}_j(N\!-\!d\!-\!j),\dots,V^{J}_j(N\!-\!j)) \ .
\eea
With this notation we have:
\bl
\label{uvPsiNPhiNdefs}
The deformation equations can be written in the folded windows  (and
dual windows) as
\be
\begin{array}{ll}
\ds{
\frac \pa{\pa u_K} \m{\Psi}_N  = {\m{\bf U}^{N,\Psi}}_{K}
\m{\Psi}_N} \ , & \qquad
\ds{ \frac \pa{\pa u_K} \m{\un\Psi}^{N\!-\!1}   =
\m{\un\Psi}^{N\!-\!1} {\m{\un{\bf  U}}^{N,\Psi}}_{K}}\ ,\\
\ds{ \frac \pa{\pa v_J} \m{\Psi}_N  = -{\m{\bf V}^{N,\Psi}}_{J}
\m{\Psi}_N}  \ , & \qquad
\ds{\frac \pa{\pa v_J} \m{\un\Psi}^{N\!-\!1}
=-\m{\un\Psi}^{N\!-\!1} {\m{\un{\bf V}}^{N,\Psi}}_{J}} \ , \label{PsiN_def}
\end{array}
\ee
\be
\begin{array}{ll}
\ds{\frac \pa{\pa u_K} \m{\Phi}_N  = -{\m{\bf U}^{N,\Phi}}_{K}
\m{\Phi}_N}  \ , & \qquad \ds{ \frac \pa{\pa u_K} \m{\un\Phi}^{N\!-\!1}   =
- \m{\un\Phi}^{N\!-\!1} {\m{\un{\bf  U}}^{N,\Phi}}_{K}}  \ , \\
\ds{ \frac \pa{\pa v_J} \m{\Phi}_N  = {\m{\bf V}^{N,\Phi}}_{J}
\m{\Phi}_N}  \ , & \qquad \ds{\frac \pa{\pa v_J} \m{\un\Phi}^{N\!-\!1}
=\m{\un\Phi}^{N\!-\!1} {\m{\un{\bf V}}^{N,\Phi}}_{J}} \ ,
\end{array}  \label{PhiN_def}
\ee
where
\be
\begin{array}{ll}
\ds{
{\m{\bf U}^{N,\Psi}}_{K} := \sum_{j=0}^K
{\m{U}^{N,d_2}}_{j,K}\A_{N\!+\!j-1}\cdots\A_N} \ ,
&
\qquad \ds{
{\m{\un {\bf U}}^{N,\Psi}}_{K} := \sum_{j=0}^K
\m{\un\B}_{N}\cdots\m{\un\B}_{N\!+\!j\!-\!1}{\m{U}^{N\!+\!d_1\!-\!1,d_1}}_{j,K}}
 \ , \\
\ds{
{\m{\bf V}^{N,\Psi}}_{J} := \sum_{j=0}^J
{\m{{ V}}^{N,d_2}}_{j,J}\A^{N\!-\!j}\cdots\A^{N\!-\!1}} \ ,
&
\qquad\ds{
{\m{\un{\bf V}}^{N,\Psi}}_{J} := \sum_{j=0}^J
 \m{\un\B}^{N\!-\!1}\cdots
\m{\un\B}^{N\!-\!j}
{\m{V}^{N\!+\!d_1\!-\!1,d_1}}_{j,J}} \ ,
\end{array}
\ee
\be
\begin{array}{ll}
\ds{
{\m{\bf U}}^{N,\Phi}_{K}
:= \sum_{j=0}^K {\m{U}^{N\!-\!j,d_1}}_{j,K}\B^{N\!-\!j}\cdots\B^{N\!-\!1}} \ ,
&
\qquad\ds{
{\m{\un{\bf U}}^{N,\Phi}}_K
:= \sum_{j=0}^K
\m{\un\A}^{N\!-\!1}\cdots\m{\un\A}^{N\!-\!j}
{\m{U}^{N\!+\!d_2\!-\!1-\!j,d_2}}_{j,K}}
\cr
\ds{
{\m{\bf V}^{N,\Phi}}_{J} := \sum_{j=0}^J
{\m{ V}^{N\!+\!j,d_1}}_{j,J}\B_{N\!+\!j\!-\!1}\cdots\B_{N}} \ ,
&
\qquad \ds{
{\m{\un{\bf V}}^{N,\Phi}}_{J} :=  \sum_{j=0}^J
\m{\un\A}_{N}\cdots\m{\un\A}_{N\!-\!1\!+\!j}
{\m{{ V}}^{N\!+\!j+\!d_2\!-\!1,d_2}}_{j,J}} \ .
\end{array}
\ee
\el

\noindent
{\bf Proof}. This follows exactly the same lines as the proof of Lemma
\ref{D12}.  \hfill  Q.E.D.

\medskip

\section{Compatibility of the finite difference--differential--deformation 
systems}

We want to now prove that the recursion relations (\ref{recs1}), 
(\ref{recs_bar1}), the linear differential systems (\ref{foursystem}) and the 
systems of deformation equations (\ref{PsiN_def}), (\ref{PhiN_def}) are all 
{\em compatible} in the sense  that they admit a {\it basis} of 
simultaneous solutions (fundamental systems).

\bp
\label{shift_dx_compatibility}
The shifts $N\mapsto N+1$ in eqs. (\ref{recs1}) implemented by $\ds{\A_N}$ and 
 $\ds{\B_N}$ and the sequence of differential equations (\ref{xDE_PsiN}),
(\ref{yDE_PhiN}), respectively, are compatible as  vector 
differential--difference systems. That is, there
 exists a sequence of $(d_2+1)\times(d_2+1)$ fundamental matrix solutions 
$\ds{\le\{\m{\bf \Psi}_{N}(x)\ri\}_{N \ge d_2+1}}$  and $(d_1+1)\times(d_1+1)$ 
fundamental matrix solutions  
$\ds{ \le\{\m{\bf \Phi}_{N}(y)\ri\}_{N\ge d_1+1}}$  simultaneously  satisfying
the equations 
\bea
&& \m{\bf \Psi}_{N\!+\!1}(x) = \A_N(x)\m{\bf \Psi}_{N}(x)\ ,\label{shiftpsi}\\
&& \frac \pa{\pa x}  \m{\bf \Psi}_{N}(x) =
 -{\m{D}^N}_1(x) \m{\bf \Psi}_{N} (x)\ , \quad N\ge d_2+1 \ ,
\label{dxpsi}
\eea
and
\bea
&& \m{\bf \Phi}_{N\!+\!1}(y) = \B_N(y)\m{\bf \Phi}_{N}(y)\ ,\label{shiftphi}\\
&& \frac \pa{\pa y}  \m{\bf \Phi}_{N}(y) 
= -{\m{D}^N}_2(y) \m{\bf \Phi}_{N} (y) \ , \quad N\ge d_1+1 \ ,
\label{dyphi}
\eea
respectively.
The same result holds for the barred quantities and  the shifts 
$N\mapsto N-1$ implemented by $\ds{\m{\un\A}^N}$ and $\ds{\m{\un\B}^N}$. That
is, there exist fundamental solutions $\{\ds{\m{\bf \un
\Psi}^{N\!-\!1}(y)}\}_{N\ge d_2\1+1 }$  of dimension $(d_2+1)\times
(d_2+1)$ and fundamental solutions  $\{\ds{\m{\bf \un\Phi}^{N\!-\!1}(x)
}\}_{N\ge d_2+1 }$ of dimension $(d_1+1)\times (d_1+1)$  simultaneously
satisfying the recursion relations and differential systems 
\bea
&& \m{\bf \un \Psi}^{N\!-\!1}(y) =\m{\bf \un \Psi}^N(y)  
\m{\un \B}^N(y)\ ,  \label{shiftpsi_bar}\\
&& \frac \pa{\pa y}  \m{\bf \un \Psi}^{N\!-\!1}(y) = \m{\bf
 \un \Psi}^{N\!-\!1} (y) {\m{\un D}^N}_2(y) \ ,  \quad N\ge d_1+1 \ ,
\label{dypsi_bar}
\eea
and
\bea
&& \m{\bf \un\Phi}^{N\!-\!1}(x)  = 
\m{\bf\un \Phi}^N(x) \m{\un\A}^N(x)\ ,\label{shiftphi_bar}\\
&& {\pa \over \pa x}  \m{\bf\un  \Phi}^{N\!-\!1}(x)  =
\m{\bf\un \Phi}^{N\!-\!1} (x) {\m{\un D}^N}_1(x) \ , 
\quad N\ge d_2+1 \ , \label{dxphi_bar}
\eea
respectively.
\ep
{\bf Proof}.  We prove the compatibility for only one of the four
shift-differential systems, the others being completely analogous.\par
 The statement amounts to proving that 
\be
\pa_x +
\m{D_1}^{N} (x) =
 \A^N \circ \le( \pa_x +\m{D_1}^{N\!+\!1}(x) \ri)\circ \A_N =  \pa_x +
\A^N(x)\m{D_1}^{N\!+\!1}(x)\A_N(x) + \A^N(x)\frac d{dx}\A_N(x)\ ,
\ee 
where the dependence of $\ds{\A_N}$ on $x$ has been emphasized.\\
Let 
\be
\m{\tilde \Psi}_N(x)
:=\le[\tilde\psi_{N\!-\!d_2}(x),\dots,\tilde\psi_N(x)\ri]^t \ , \quad N\geq d_2
\ee
be {\em any} solution to the equation  
\be
  \le(\pa_x +
\m{D_1}^{N} (x)\ri) \m{\tilde\Psi}_N(x) = 0\ .
\ee
At this stage the labeling $N-d_2,\dots,N$ has no particular meaning
because there are no $\tilde\psi_n(x)$'s with $n\not\in [N-d_2,N]$.
Nevertheless we can {\em define} 
\be \ds{ \m{\tilde\Psi}_{N\!+\!j} (x):= \A_{N\!+\!j\!-\!1}\cdots\A_N
\m{\tilde\Psi}_N (x)} \, \quad j\geq 1 \  ,
\ee
and 
\be \ds{ \m{\tilde\Psi}_{N\!-\!j} (x):= \A^{N\!-\!j}\cdots\A^{N\!-\!1}
\m{\tilde\Psi}_N (x)} \ , \quad 0\leq j \leq N-d_2
\ee
Because of the recursive structure of the matrices $\ds{\A_N}$, the above
definition, e.g., for $\ds{\m{\tilde \Psi}_{N\!+\!1}}$ actually
defines not $d_2+1$ new functions, but only one new function:
 $\tilde\psi_{N\!+\!1}$. Therefore, componentwise, 
 we have defined a sequence of new
functions $\tilde \psi_{m}(x)$, which satisfy the recursion relation 
\be
x\tilde \psi_{m}(x)  = \sum_{j=-1}^{d_2} \a_j(m)\tilde \psi_{m-j}(x)\ ,
m\geq d_2  \ . \label{tilderecursion}
\ee
By this definition and by the structure of the matrix 
\be
\m{D_1}^N = \stackrel{N}{\gamma}\m{\A}^{N-1}  + 
 \stackrel{N}{\b_0}  +  \sum_{j=1}^{d_1}\stackrel{N}{\b_j} {\A_{N+j-1}}
{\A_{N+j-2}}\cdots {\A_N}  \ ,
\ee
the differential system componentwise now reads 
\be
\pa_x\tilde \psi_{n}(x)  = -\sum_{j=-1}^{d_1} \b_j(n+j)\tilde \psi_{n+j}(x)
\ ,\qquad n=N-d_2, \dots , N\ ,  \label{tildediffrecursion}
\ee
where the $\tilde \psi_n$'s that fall outside the window $N-d_2,\dots,N$ have
been defined above in terms of the ones within the window. 
Therefore we need to prove that the newly defined  function
\be 
\tilde \psi_{N+1}(x) :=
\frac {x-\a_0(N)}{\gamma(N)}\tilde  \psi_N(x) -\sum_{l=1}^{d_2} 
\frac {\a_l(N)}{\gamma(N)}\tilde \psi_{N-l}(x)
\ee
 satisfies the same sort of differential equation as the preceding ones.
(A simple argument by induction then
shows that all $\tilde \psi_{N+j}$ satisfy the same sort of differential
equation for any $j>1$).
This in turn amounts to proving that 
\be
\pa_x\tilde \psi_{N+1}(x) =  -\sum_{j=-1}^{d_1} \b_j(N+1+j)\tilde
 \psi_{N+1+j}(x)\label{LHS1}\ . 
\ee
To do this we compute 
\bea
 {\gamma(N)} \pa_x\tilde\psi_{N+1}(x)  
=&&\frac d{dx}
\le( x\tilde\psi_N(x)-\sum_{l=0}^{d_2}\a_l(N)\tilde\psi_{N-l}(x) \ri)
 \cr
=&&
\sum_{l=0}^{d_2}\sum_{j=-1}^{d_1}\a_l(N)\b_j(N\!-l\!+\!j)
\tilde\psi_{N\!-l\!+\!j}(x) + \tilde\psi_N(x) -x\sum_{j=-1}^{d_1}\b_j(N+j)
\tilde\psi_{N+j}(x) = \cr
=&& \sum_{l=0}^{d_2}\sum_{j=-1}^{d_1}\a_l(N)\b_j(N\!-l\!+\!j)
\tilde\psi_{N\!-l\!+\!j}(x) +\tilde  \psi_N(x)  \cr
&& -\sum_{l=-1}^{d_2}\sum_{j=-1}^{d_1}\a_l(N+j)\b_j(N+j) 
\tilde \psi_{N+j-l}(x)\label{RHS1}\ .
\eea
Rearranging eqs. (\ref{LHS1}, \ref{RHS1}), we have to prove the
identity
\bea
 -\gamma(N)\sum_{j=-1}^{d_1} \b_j(N+1+j)\tilde \psi_{N+1+j}(x) 
= && \sum_{l=0}^{d_2}\sum_{j=-1}^{d_1}\a_l(N)\b_j(N\!-l\!+\!j) 
\tilde\psi_{N\!-l\!+\!j}(x) + \tilde \psi_N(x)  \cr
&& -\sum_{l=-1}^{d_2}\sum_{j=-1}^{d_1}\a_l(N+j)\b_j(N+j) 
\tilde\psi_{N+j-l}(x) \ ,
\eea
or equivalently 
\bea
  \tilde \psi_N(x) =  &&-
\sum_{l=-1}^{d_2}\sum_{j=-1}^{d_1}\a_l(N)\b_j(N\!-l\!+\!j)
\tilde\psi_{N\!-l\!+\!j}(x)  \cr
&& +\sum_{l=-1}^{d_2}\sum_{j=-1}^{d_1}\a_l(N+j)\b_j(N+j) 
\tilde \psi_{N+j-l}(x)\ .\label{ddd}
\eea
But this last equation is nothing but the Heisenberg commutation relations  
in eq. (\ref{heisenbergone}) and eq. (\ref{heisenbergtwo}).  This means that
rearranging the coefficients in  front of $\tilde\psi_{N+r}(x)$ in the RHS of
eq. (\ref{ddd}), the only nonvanishing coefficient is that of
$\tilde\psi_N(x)$ and it is exactly $1$. A similar argument may be used to
prove that the relations  (\ref{tildediffrecursion}) hold also for $1\leq n <
N-d_2$

For future convenience we remark that this verification amounts to the fact
that the coefficients of the $\tilde\psi_{N+r}(x)$'s are the same as for the 
orthogonal quasi-polynomials $\psi_{N+r}(x)$, since it relies only on 
the recursion relations,  which are the same. Given that the quasi-polynomials 
$\psi_{N+r}(x)$ are linearly independent, the equality of LHS and RHS follows 
also for any other sequence. \hfill Q.E.D.
\vskip 3pt

Prop. \ref{shift_dx_compatibility} means that we can define $d_2+1$
sequences of functions $\le\{\psi^{(q)}_n(x)\ri\}_{n\in N, q=0 \dots d_2}$ in
such  a way that in any ``window'' of size $d_2+1$ they constitute a
fundamental  system of solutions to the differential system (\ref{dxpsi}).  One
of these  sequences is obviously provided by the orthogonal quasi--polynomials.
Each of  them satisfies both the recursion relations 
\be
x\psi_n^{(q)} (x) = \sum_{l=-1}^{d_2}\a_l(n)\psi_{n-l}^{(q)}(x)
\ee
and the derivative relations
\be
\pa_x\psi_n^{(q)}(x) = - \sum_{j=-1}^{d_1} \b_j(n+j)\psi_{n+j}^{(q)}(x)\ . 
\ee
\br
\label{remark1}
In principle, in order to define these $d_2+1$ sequences one should
solve the differential system (\ref{dxpsi}) in a given window and then define 
recursively the rest of the sequence backwards and forwards.
To pass from the semi-infinite case to the infinite one, we may define
the full sequence $\psi_n^{(q)}$ for $n\in \mathbb Z$ just by application of
products of the matrices $\ds{\A_N}$ and their inverses, provided the 
$\alpha_j(n)$'s are so defined that all the $\ds{\A_N}$'s are invertible.
\er
In a completely parallel manner we can define $d_1+1$ sequences of
functions  $\le\{\phi^{(q)}_n(y)\ri\}_{n\in N, q=0 \dots d_1}$
which provide fundamental systems satisfying
\be
 \le[\pa_y +\m{D_2}^N(y)\ri]\m{\bf \Phi}_N(y)= 0\ .
\ee
Moreover, with minor modifications, we can construct analogous sequences
for the dual systems 
\bea
\pa_y\m{\bf\un \Psi}^{N\!-\!1} (y) & = &\m{\bf\un
\Psi}^{N\!-\!1}(y)\m{\un  
D_2}^N(y)\\
 \pa_x\m{\bf \un \Phi}^{N\!-\!1}(x) & = & 
\m{\bf \un \Phi}^{N\!-\!1}(x) \m{\un D_1}^N(x)\ .
\eea
 The only difference is that the matrices $\ds{\m{\un{\bf a}}^N}$ and
$\ds{\m{\un{\bf b}}^N}$  now implement the shift $N\to N-1$. The barred
sequences  $\le\{\un\phi^{(q)}_n(y)\ri\}_{n\in \bbN, q=0..d_1}$ and
$\le\{\un\psi^{(q)}_n(y)\ri\}_{n\in \bbN, q=0..d_1}$
 will therefore satisfy the recursion relations 
\bea
x\un\phi^{(q)}_n(x) &= {\ds
\sum_{l=-1}^{d_2} \alpha_{l}(n+l)\un\phi^{(q)}_{n+l}(x)\ ,
\quad \pa_x \un\phi^{(q)}_n(x)} &= \ds{
\sum_{j=-1}^{d_1} \b_{j}(n)\un\phi^{(q)}_{n-j}(x)}\ ,\\
y\un\psi^{(q)}_n(y) &= {\ds \sum_{j=-1}^{d_1}
\b_{j}(n+j)\un\psi^{(q)}_{n+j}(x)\ , \quad  \pa_y \un\psi^{(q)}_n(y)} &= 
\ds{\sum_{l=-1}^{d_2} \a_{l}(n)\un\phi^{(q)}_{n-l}(y)}\ . 
\eea
A completely analogous statement holds for the matrices defining the
deformation equations in any window.
\bp
\label{shift_du_compatibility}
The shifts $N\mapsto N+1$ implemented by $\ds{\A_N}$ in eq.(\ref{shiftpsi})
and  the sequence of differential equations
\be
 \pa_{u_K}\m{\bf \Psi}_N = {\m{\bf U}^{N,\Psi}}_{K}\m{\bf \Psi}_N ,
\qquad \pa_{v_J}\m{\bf \Psi}_N =
-{\m{\bf V}^{N,\Psi}}_{J}\m{\bf \Psi}_N , \qquad N\geq d_2+1 \label{defUVPsi}
\ee
 are compatible, as are the shifts $N\mapsto N-1$ implemented by $\ds{\m{\un
{\bf b}}^N}$ in (\ref{shiftpsi_bar}) and the sequence of differential equations
 \be  
\pa_{u_K}\m{\un{\bf \Psi}}^{N\!-\!1} =
\m{\un{\bf \Psi}}^{N\!-\!1}{\m {\un{\bf U}}^{N,\Psi}}_{K}  ,\qquad  
\pa_{v_J}\m{\un{\bf \Psi}}^{N\!-\!1} =
-\m{\un{\bf \Psi}}^{N\!-\!1}{\m {\un{\bf V}}^{N,\Psi}}_{J} \ , 
\qquad N\geq 1 \ .
\label{defUVPsi_bar}
\ee
Similarly the shifts implemented by $\ds{\B_N}$ and $\ds{\m{\un{\bf a}}^N}$ 
are  compatible with the equations
\bea
\pa_{u_K}\m{\bf \Phi}_N &=&
-{\m {\bf U}^{N,\Phi}}_{K}\m{\bf \Phi}_N \ , \qquad
 \pa_{v_J}\m{\bf \Phi}_N =
{\m {\bf V}^{N,\Phi}}_{J}\m{\bf \Phi}_N \ , \qquad N\geq d_1+1  
\label{defUVPhi}\\
\pa_{u_K}\m{\un{\bf \Phi}}^{N\!-\!1} &=&
-\m{\un{\bf \Phi}}^{N\!-\!1}{\m {\un{\bf U}}^{N,\Phi}}_K   \ , \qquad
\pa_{v_J}\m{\un{\bf \Phi}}^{N\!-\!1} =
\m{\un{\bf \Phi}}^{N\!-\!1}{\m {\un{\bf V}}^{N,\Phi}}_J  \ , \qquad 
N\geq 1\ .  \label{defUVPhi_bar}
\eea
\ep
{\bf Proof}.
We will prove compatibility of only one of the eight kinds of systems
with the shift; the remaining cases are proven similarly.\par
As for Prop. \ref{shift_dx_compatibility}, we first define a continuous
parametric family (depending on $x$) of solutions to the system 
\be
\le[\pa_{u_K} -{\m{\bf U}^{N,\Psi}}_{K}\ri]\m{\tilde \Psi}_N =0\ .
\ee
We then define the shifted functions 
 $\ds{
\m{\tilde\Psi}_{N\!+\!j}:=
 \A_{N\!+\!j\!-\!1}\cdots\A_N
\m{\tilde\Psi}_N }$ so that the equation reads, componentwise
\be
\pa_{u_K}\tilde \psi_n = \sum_{j=0}^{K} U^{K}_j(n)\tilde\psi_{n+j}\ ,\qquad
n=N-d_2,\dots,N\ .
\ee
Then we have to check that the newly defined $\tilde\psi_{N+1}$ also
satisfies 
\be
 \pa_{u_K}\tilde\psi_{N\!+\!1}  = \sum_{j=0}^{K}
U^{K}_j(N\!\!+\!1)\tilde \psi_{N\!+\!1\!+\!j}
\ee 
(and by induction the corresponding equations for 
$\tilde \psi_{N+r},\ r\geq 1$).
As before, we use the relations
\be
\tilde\psi_{N\!+\!1} =
\frac {x-\a_0(N)}{\gamma(N)} \tilde\psi_N(x) -\sum_{l=1}^{d_2} 
\frac {\a_l(N)}{\gamma(N)}\tilde\psi_{N-l}(x)
\ee
  and the differential
system satisfied by the $\tilde\psi_{N-d_2},\dots,\tilde\psi_{N}$.
To conclude the equality we can reason as in the remark in the proof of 
Prop. \ref{shift_dx_compatibility}.  \hfill Q.E.D.

The final proposition in this section assures that the deformation equations 
are compatible with the $x,y$ differential equations in all windows.
\bp
\label{dx_du_compatibility}
The system of equations 
\bea
\le(\pa_x+\m {D_1}^N\ri)\m{\bf \Psi}_{N}(x) &=&0   \ , \label{eqdxpsi}\\ 
\le(\pa_{u_K} - {\m{\bf U}^{N,\Psi}}_{K}\ri) \m{\bf \Psi}_{N}(x) &=& 0  \ ,
\\
\le(\pa_{v_J} + {\m{\bf V}^{N,\Psi}}_{(J)}\ri)\m{\bf \Psi}_{N}(x) &=& 0  \ ,
 \label{eqsPsiNvjx}\\
\m{\bf \Psi}_{N\!+\!1}(x) &=& \A_N(x) \m{\bf \Psi}_{N}(X) \label{eqshiftpsi} 
\ ,  \quad N\ge d_2+1 \ ,
\eea
is compatible, and hence sequences of fundamental systems of solutions 
$\ds{ \le\{\m{\bf \Psi}_{N}(x)\ri\}_{N > d_2}}$ to all equations exist.
The same statement holds for the system
\bea
\le(\pa_y+\m {D_2}^N\ri)\m{\bf \Phi}_{N}(y) &=&0  \label{eqdyphi} \ ,\\ 
\le(\pa_{u_K} + {\m{\bf U}^{N,\Phi}}_{K}\ri) \m{\bf \Phi}_{N}(y) &=& 0 \ ,
\\
\le(\pa_{v_J} - {\m{\bf V}^{N,\Phi}}_{(J)}\ri)\m{\bf \Phi}_{N}(y)&=& 0 \ ,
 \label{eqsPhiNvjy}\\
\m{\bf \Phi}_{N\!+\!1}(y) &=& \B_N(x) \m{\bf \Phi}_{N}(y) \ ,
 \quad N\ge d_1+1 \ .
\label{eqshiftphi}
\eea
The corresponding systems for the barred sequences are also compatble, and
hence also admit simultaneous sequences of fundamental solutions 
${\ds \m{\bf \un\Psi}^{N-1}}(x)$ and 
${\ds \m{\bf \un \Phi}^{N-1}}(y)$.
\ep
{\bf Proof}.
The proof will only be given for the system (\ref{eqdxpsi})-(\ref{eqshiftpsi})
since the others are proved in the same way. The compatibility follows from
Props.  \ref{shift_dx_compatibility} and \ref{shift_du_compatibility} together
with a proof of compatibility of the equations
(\ref{eqdxpsi})-(\ref{eqsPsiNvjx}). Indeed, from the $d_2+1$ functions in 
\be
\m{\tilde \Psi}_N =
[\tilde\psi_{N\!-\!d_2}(x,u,v),\dots,\tilde\psi_N(x,u,v)]^t  
\ee
we can consistently define a whole sequence of functions $\tilde\psi_n$'s by
means of the $x-$recursion relations in such a way that componentwise the 
system reads 
\bea
\pa_x\tilde\psi_n &=& -\sum_{j=-1}^{d_1}\b_j(n+j) \tilde\psi_{n+j} \ , 
\label{psindxtilde}\\
\pa_{u_K} \tilde\psi_n &=& \sum_{j=0}^{K}U^K_j(n)\tilde \psi_{n+j} \ ,
\label{psinduktilde} \\
\pa_{v_J} \tilde\psi_n &=& -\sum_{j=0}^{J}V^J_j(n-j)\tilde \psi_{n-j} \ .
\label{psindvjtilde}
\eea
Taking cross derivatives and using these expressions one gets, e.g.,
\bea
&&
\pa_{u_k}\pa_x\tilde \psi_n = -\sum_{k=0}^K\sum_{j=-1}^{d_1}U^K_k(n+j)
\b_j(n+j)\tilde \psi_{n+j+k} - \sum_{j=-1}^{d_1}
\le(\frac \pa{\pa u_K}\beta_j(n+j)\ri)\tilde \psi_{n+j} \ , \label{or}
\\
&&
\pa_x\pa_{u_k}\tilde \psi_n = -\sum_{j=-1}^{d_1} \sum_{k=0}^K U^K_k(n)
\b_j(n+j+k)\tilde \psi_{n+j+k}\ .
\eea
The expressions for the derivatives of the coefficients $\beta_j$
may be obtained from the deformation equation (\ref{deformPu}) for $P$
and then substituted  back into eq. (\ref{or}). However, to prove the equality
of the two cross derivatives it is sufficient to collect the coefficients of
$\tilde \psi_{n+q}$ and note that exactly the same coefficients appear
when the functions $\{\tilde\psi_n\}$ are replaced by the
orthogonal quasi-polynomials $\{\psi_n\}$, for which the equality of the
two expressions certainly holds. Since the orthogonal quasi-polynomials are
linearly  independent functions, the individual coeficients must agree.
(This is essentially the same argument as in the remark at the end of proof of
Prop. \ref{shift_dx_compatibility}).
The mutual compatibility of the $\pa_{u_K}$ and $\pa_{v_{J}}$
deformations is proved in exactly the same way. One just takes the
 the cross derivatives in equations (\ref{psinduktilde}) and
(\ref{psindvjtilde}) and notes that, since these are the same as for the 
case of the orthogonal quasi-polynomials $\{\psi_n\}$, the
corresponding coefficients in the cross differentiated expression must
 be equal. 
\hfill  Q.E.D.
\medskip

\section{Spectral Duality}

The aim of this section is to state and prove some remarkable
relations between systems related to dual windows (in the sense of
Def. \ref{dualwindows}), which will justify the terminology.
One of the main results will be that  the four spectral curves
 given by the characteristic polynomials of  $D_1,\un D_1,D_2,\un D_2$
associated to  the  four systems (eq. (\ref{foursystem}))
 on the two pairs of dual windows $\ds{(\m{\Psi}_N,\m{\un
\Phi}^{N\!-\!1})   }$,  $\ds{(\m{\Phi}_N,\m{\un
\Psi}^{N\!-\!1} )  }$ are actually
the same curve.

\subsection{Dual spectral curves}

First we need a linear algebra lemma.
\bl 
\label{tensor}
Let $T$ be a square  matrix having  the block form 
\be
T = \le[\begin{array}{c|c|c|c|c}
0 & F_1 & 0 & 0 &0 \\[2pt]
\hline
0&0&F_2 & 0 & 0\\[2pt]
\hline
0&0&0& \ddots  &0\\
\hline
0&0&0&0&F_d\\[2pt]\hline
G_0&G_1&G_2&\cdots &G_d
\end{array}
\ri]\ ,
\ee
where the $d+1$ blocks have compatible sizes and the diagonal blocks
are square. Then 
\bea
&&\det\le[\1 - T\ri] = \det\le[ \1 - D \ri]\ ,\hbox { where}\\
&& D:= G_d + \sum_{k=0}^{d-1} G_k \cdot F_{k\!+\!1 } \cdots F_{d}\ ,
\eea
and $\1$ denotes, according to the context, the unit matrix of
appropriate size.
\el
{\bf Proof}.
Let 
\be
N_F:= \le[\begin{array}{c|c|c|c|c}
0 & F_1 & 0 & 0 &0 \cr
\hline
0&0&F_2 & 0 & 0\cr
\hline
0&0&0&\ddots &0\cr
\hline
0&0&0&0&F_d\cr\hline
0&0&0&0&0
\end{array}
\ri]\ .
\ee
We multiply the  matrix $\1-T$ from the right by the matrix
\be
\le(\1-N_F\ri)^{-1}= \le[\begin{array}{c|c|c|c|c}
\1  & F_1 & F_1F_2 & F_1F_2F_3 &F_1\cdots F_d \cr
\hline
0&\1&F_2 & F_2F_3 & F_2\cdots F_d\cr
\hline
0&0&\ddots&\ddots &\vdots\cr
\hline
0&0&0&\1&F_d\cr\hline
0&0&0&0&\1
\end{array}
\ri]
\ee
Since the matrix $(\1-N_F)^{-1}$ is unimodular, the determinant of
$\1-T$ remains unaffected. Then one computes
\be
(\1-T)\cdot(\1-N_F)^{-1} = \le[\begin{array}{c|c|c|c|c}
\1  & 0 & 0 & 0 &0 \cr
\hline
0&\1&0 & 0 & 0\cr
\hline
0&0&\1&0 &0\cr
\hline
0&0&0&\ddots&0\cr\hline
\star&\star&\star&\star &\1 - D
\end{array}
\ri] \ ,
\ee
from which the statement follows by taking the determinant.  \hfill Q.E.D.

\smallskip

\bp
\label{accordion}
The spectral curves  associated to  the characteristic polynomials of
$\ds{\m{D_1}^N,\ \m{\un D_2}^N,\ \m{D_2}^N,\ \m{\un D_1}^N}$
are pairwise equal. More precisely, we have the formulae
\bea
&& u_{d_1\!+\! 1}\det\le[x\1_{d_1+1 }-{\m{D}^N}_2 (y) \ri]  =
v_{d_2\!+\! 1}\det\le[y\1_{d_2+1} - {\m{\un D}^N}_1 (x)\ri] \ ,
 \label{dual1} \\ 
&& u_{d_1\!+\! 1} \det\le[x\1_{d_1+1 }-{\m{\un D}^N}_2(y) \ri]  =
v_{d_2\!+\! 1} \det\le[y\1_{d_2+1} -  {\m{D}^N}_1 (x)\ri]\ , \label{dual2}
\eea
which connect the spectral curves of the differential operators of different 
dimensions operating on the two pairs of dual windows.
\ep
{\bf Proof}. 
We will only prove one equality since the other is  proved similarly.
We start with the computation of the characteristic polynomial of
$D_1(x)$ 
 \bea
\det\le[
y\1-\m{D_1}^N(x) \ri] &=&
\det\le[ -\stackrel{N}{\gamma} \m{\A}^{N-1}
 \ri] \det\le[ \1 - {\A_{N-1}}\m{\gamma}_{N}\le(y\1 - \stackrel{N}{\b_0}
- \sum_{j=1}^{d_1} \stackrel{N}{\b_j} {\A_{N\!+\!j\!-\!1}\cdots \A_N} \ri)
  \ri] \cr
&=& \frac
{\gamma(N-1)^2}{v_{d_2+1}} \det\le[ \1 -
{\A_{N-1}}\m{\gamma}_{N} \le(y\1 - \stackrel{N}{\b_0}
- \sum_{j=1}^{d_1} \stackrel{N}{\b_j} {\A_{N\!+\!j\!-\!1}\cdots\A_ N} \ri)
 \ri]\ ,
\eea
where we have used the identity in eq. (\ref{integratedCCR}).
Now we use Lemma \ref{tensor} with the identifications 
\be
G_{d_1} =
{\A_{N-1}}\m{\gamma}_{N} \le(y\1 - \stackrel{N}{\b_0}\ri)\ , \quad
G_k =  -{\A_{N-1}}\m{\gamma}_{N}{\m{\b}^{N}}_{d_1\!-\!k\!-\!1}
\ , \ k=1\dots d_1, \quad  F_k = \m{\A}_{N\!+\!d_1\!-\!k}\ ,
\ee
 and thus obtain
\bea 
&& \det\le[ \1 -{\A_{N-1}} \m{\gamma}_{N} \le(y\1 -
\stackrel{N}{\b_0} - \sum_{j=1}^{d_1}\stackrel{N}{\b_j}
{\A_{N\!+\!j\!-\!1} \cdots\A_N} \ri)
 \ri] =
\det\le[\1_{(d_1+1)(d_2+1)} - T_{\A\B}\ri]\ .
\eea
The matrix  $T_{\A\B}$  is defined by
\bea
 T_{\A\B}&&:=\le[
\begin{array}{c|c|c|c}
0&\ds{\A_{N\!+\!d_1\!-\!1}}  & 0 & 0\\ [7pt]
\hline
0&0 & \ddots  & 0 \\[7pt]
\hline
0&0 & 0 & \ds{\A_{N}} \\[7pt]
\hline
\ds{\!\!-\!\!\!\! {\A_{N-1}}
\!\!\stackrel{N}{\b_{d_1}} \!\!\m{\gamma}_N\!\!} &  
\ds{\!\!-\!\!\!\!{\A_{N-1}} 
\!\!\stackrel{N}{\b_{d_1\!-\!1}}\!\!\m{\gamma}_N\!\!} & 
\cdots &
\ds{{\A_{N-1}} \!\!(y\1-\stackrel{N}{\b_0})\!
\m{\gamma}_N\!\! }
\end{array} \ri]  \\
&&
 =  \le[
\begin{array}{c|c|c|c}
\ds{\!\!\A_{N\!+\!d_1\!-\!1}\!\!}&  0 & 0 & 0 \\ 
\hline
0 &\ddots & 0 & 0 \\
\hline
 0 & 0 &\ds{\!\!\A_{N}\!\!} & 0  \\ 
\hline
 0 & 0 & 0 &\ds{\!\!\A_{N-1}\!\!} 
\end{array} \ri]
\le[
\begin{array}{c|c|c|c}
 0 & \1  &0 & 0 \\[3pt]\hline
 0 & 0 & \ddots   & 0\\[3pt]\hline
 0 &0 & 0 & \!\!\1 \\[3pt]\hline
\ds{\!\!-
\!\!\stackrel{N}{\b_{d_1}} \!\!\m{\gamma}_N\!\!} &  
\ds{\!\!-\!\!\stackrel{N}{\b_{d_1\!-\!1}}\!\!\m{\gamma}_N\!\!} & 
\cdots &
\ds{ \!\!(y\1-\stackrel{N}{\b_0})\!\!
\m{\gamma}_N\!\! }
\end{array} \ri] \ , 
\eea
We regard $T_{\A\B}$  as an endomorphism of
$\C^{d_2+1}\otimes\C^{d_1+1}$. 
 Let  $P_{12}$ be the involution interchanging the two
factors of the tensor product 
 and let the matrix $C$  implement the reversal of  order endomorphism within 
the $(d_2+1)\times (d_2+1)$ blocks: 
\be
C:= {\rm Blockdiag}\overbrace{( R,R,..,R)}^{d_2+1 \hbox{ times}}\ , \
\ R:=\le[
\begin{array}{ccccc}
0&0&0&0&1\cr
0&0&0&1&0\cr
0&0&\cdot^{\ds{\cdot^{\ds{\cdot}}}}&0&0\cr
0&1&0&0&0\cr
1&0&0&0&0
\end{array}
\ri]\in GL_{d_1+1} \ .
\ee
A direct inspection shows  
\bea
C P_{12}T_{\A\B}^t P_{12}C^{-1} = \le[
\begin{array}{c|c|c|c|c}
0& \stackrel{N\!-\!d_2}{\un\B} & 0 &0 &0\\[7pt] \hline
 0 &0&\ddots  & 0 & 0\\[7pt]\hline
 0&0 &0& \stackrel{N-2}{\un\B} & 0 \\[7pt]\hline
0& 0 &0 & 0 &
\stackrel{N\!-\! 1} {\un\B}\\[7pt]\hline
\ds{\!\!-\!\!\m{\un \B}^{N}
\m{\underline \a_{d_2}}^{N-1}\!\!\!\m{\underline
\gamma}_{N-1}  \!\!} &
\ds{\!\!-\!\! \m{\un\B}^{N} 
\m{\underline \a_{d_2\!-\!1}\!\!}^{N-1}\!\!\!\m{\underline
\gamma}_{N-1} \!\!} &
\cdots &
\ds{\!\!-\!\!\m{\un \B}^{N} \!\!
\m{\underline \a_1}^{N-1}\!\!\m{\underline
\gamma}_{N-1}\!\! } & 
\ds{ \m{\un\B}^{N}
\le(x\1\!-\!\m{\underline \a_0}^{N-1}\!\ri)\!\m{\underline
\gamma}_{N-1}  }
\end{array}
\ri]
\eea
where we  now have $d_2+1$ square blocks of dimension $d_1+1$ (i.e. the
number of blocks and the dimensions of the blocks have been
interchanged).
 The barred symbols are precisely those defined in eqs. (\ref{abarN}, 
\ref{bbarN}), (\ref{alphabarjN},  \ref{betabarjN}).

We now use  Lemma \ref{tensor} again to get \bea
\frac {v_{d_2+1}}{\gamma(N-1)^2}\,\det\le[y\1 - {\m{
D}^{N}}_1(x)\ri] &=& \det\le[\1_{(d_1+1)(d_2+1)}-T_{\A\B}\ri] \cr
&=&
\det\le[\1_{(d_1+1)(d_2+1)}-C P_{12}T_{\A\B}^t P_{12}C^{-1}\ri] \cr 
&=&
\det\le[\1_{d_1+1} - \m{\underline\gamma}_{N-1}
\m{\un \B}^{N} \le( x\1-\m{\underline\a_0}^{N-1}
-\sum_{j=1}^{d_2} \m{\un\B}^{N-1}\cdots \m{\un \B}^{N-j}
\m{\underline\a_j}^{N-1}\ri)\ri] \cr
&=&
\det\le[-\m{\un\gamma}_{N-1}\m{\un\B}^{N} \ri]\,\det\le[x\1 - {\m{\un
D}^{N}}_2(y)\ri] \cr
&=& \frac {u_{d_1+1}}{\gamma(N-1)^2} \,\det\le[x\1 - {\m{\un
D}^{N}}_2(y)\ri]\ .
\eea
This concludes the proof.  \hfill Q.E.D.

\br
Notice that this proof was based purely on an algebraic reinterpretation
of the characteristic equations for $\m{D_1}(x)$ and $\m{\un D_2}(y)$
in which the same set of recursion parameters $\{\a_j(n)$, $\b_j(n)\}$
appear. No assumption was required about any relations between these 
parameters, and therefore the equalities (\ref{dual1}), (\ref{dual2})
are really just identities.
\er

\subsection{Duality pairings}

In what follows we will derive a deeper form of duality; namely, that the
linear differential equations satisfied by dual windows are also dual, in
the sense of having the same associated spectral curves. This follows from
the generalized Christoffel--Darboux formulae satisfied by kernels $K_{ij},
i,j = 1, 2$ when the biorthogonal polynomials and their Fourier-Laplace
transforms are replaced by {\em any} solution to the equations of
Prop. \ref{dx_du_compatibility}.
\bp
If  $\{\un{\tilde\psi}_n(y)\}_{n\in \N}$ and $\{\tilde\phi_n(y)\}_{n\in\N}$ are
two arbitrary sequences of functions satisfying both the recursion relations 
under multiplication by $y$ and the differential relations under application
of  $\pa_y$ (constructed as in Prop. \ref{shift_dx_compatibility}), then
\be
\le(\pa_{y'}+\pa_{y}\ri) 
 \le(\m{\tilde{\un\Psi}}^{N\!-\!1}(y'),\Bmat^N\m{\tilde\Phi}_N (y)\ri) = 
\le(y -y'\ri) \le(\m{
\un{\tilde\Psi}}_N (y'), (\Amat^N)^t\m{\tilde \Phi}^{N\!-\!1}(y)\ri)\ .
\label{dyplusdyprimePsiPhi}
\ee
and
\be
\le(\pa_{x'}+\pa_{x}\ri) 
 \le(\m{\tilde{\un\Phi}}^{N\!-\!1}(x'),\Amat^N\m{\tilde\Psi}_N (x)\ri) = 
\le(x -x'\ri)\le(\m{
\un{\tilde\Phi}}_N (x'), (\Bmat^N)^t\m{\tilde \Psi}^{N\!-\!1}(x)\ri)\ .
\label{dxplusdxprimePhiPsi}
\ee
\label{mainlemma}
\ep
{\bf Proof}. We shall prove the equality  (\ref{dyplusdyprimePsiPhi}) only;
since (\ref{dxplusdxprimePhiPsi}) is proved identically.   The expressions
on either side of eq. (\ref{dyplusdyprimePsiPhi}) read (understanding the
$\un{\tilde\psi}_n$'s  to depend on $y'$ and the  $\tilde\phi_n$'s on $y$):
\bea
&&\le(\pa_{y'} + \pa_y\ri) 
 \le(\m{\un{\tilde\Psi}}^{N\!-\!1}(y'),\Bmat^N \m{\tilde \Phi}_N (y)\ri) \cr
 &=& \le(\pa_{y'} + \pa_y\ri) \le(
 -\gamma(N-1)\un{\tilde\psi}_{N-1} \tilde\phi_{N} + \sum_{j=1}^{d_1}
\sum_{k=0}^{j-1} \beta_{j}(N+k) \un {\tilde\psi}_{N+k} \tilde\phi_{N-j+k}\ri) 
 \cr
&=&\gamma(N\!-\!1) \,\un{\tilde\psi}_{N\!-\!1}\sum_{l=-1}^{d_2}
\a_l(N\!+\!l)\tilde \phi_{N\!+ l} - \gamma(N\!-\!1)
\sum_{l=-1}^{d_2}\a_l(N\!-\!1) \un{\tilde\psi}_{N\!-\!1\!-l} \,\tilde \phi_{N}
 \cr
&&  + \sum_{j=1}^{d_1}\sum_{k=0}^{j-1}\sum_{\!l=\!-\!1}^{d_2}
\a_l(N\!+\!k)\b_j(N\!  +\!k)
\un{\tilde\psi}_{N\!+\!k\!-l}\,\tilde\phi_{N\!+\!k\!-\!j} \cr
&& - \sum_{j=1}^{d_1}\sum_{k=0}^{j-1}\sum_{\!l=\!-\!1}^{d_2}
\b_j(N\!+\!k) \a_l(N\!+\!k\!-\!j\!+\!l) \un{\tilde\psi}_{N\!+\!k}
\, \tilde\phi_{N\!+\!k\!-\!j\!+l} 
\label{LHS}
\eea
and
\bea
&&\le(y' -  y\ri) \le(\m{
\un{\tilde\Psi}}_N (y'),\Amat^N\m{\tilde\Phi}^{N\!-\!1}(y)\ri)  \cr
&=& (y'-y)\le(
\gamma(N-1) \un{\tilde\psi}_N\tilde\phi_{N-1} -
\sum_{j=1}^{d_2} \sum_{k=0}^{j-1} \a_j(N+k) 
\un{\tilde\psi}_{N+k-j} \tilde\phi_{N+k} \ri)  \cr
 &=&\gamma(N\!-\!1)
\,\tilde\phi_{N\!-\!1}\sum_{j=-1}^{d_1}
\b_j(N\!+\!j) \un{\tilde\psi}_{N\!+\!j} - \gamma(N\!-\!1) \,\un{\tilde\psi}_{N}
\sum_{j=-1}^{d_1}\b_j(n\!-\!1)\tilde\phi_{N\!-\!1\!-\!j}   \cr
&&  - \sum_{l=1}^{d_2}\sum_{k=0}^{l-1}\sum_{\!j=\!-\!1}^{d_1}
\a_l(N\!+\!k)
\b_j(N\!+\!k\!-\!l\!+\!j)\un{\tilde\psi}_{N\!+\!k\!-\!l\!+\!j}\,
\tilde\phi_{N\!+\!k}  \cr
&& +  \sum_{l=1}^{d_2}\sum_{k=0}^{l-1}\sum_{\!j=\!-\!1}^{d_1}
\a_l(N\!+\!k)\b_j(N\!+\!k) \un{\tilde\psi}_{N\!+\!k\!-\!l} 
\tilde \phi_{N\!+\!k\!-\!j}\ .
\label{RHS}
\eea
\hspace{0.4cm} The claim now is that these two expressions are the same. There 
are two ways of proving this.
The first is a straightforward computation collecting all bilinear
terms of the form 
\be
F_{pq}(y',y) := \un{\tilde\psi}_{N+p}(y')\tilde
\phi_{N+q}(y)
\ee  
in the difference between (\ref{LHS}) and (\ref{RHS}) and proving that their
coefficients vanish.  The coefficient of $F_{pq}$ with $p\geq 0$ and $q\geq 0$
vanish identically, as do the coefficient of  $F_{pq}$ with $p<-1$ and $q<-1$.
The coefficient of $F_{pq}$ with $p\geq 0$ and $q<-1$ vanishes due to relation
(\ref{heisenbergone}) with $l=p-q-1$ and $n=N+q$.
The coefficient of $F_{pq}$ with $p<-1$ and $q\geq 0$ vanishes due to relation
(\ref{heisenbergtwo}) with $l=q-p-1$ and $n=N+p$.
These  cancellations are summarized in Table \ref{table}.

\begin{table}[bt]
\begin{small}
$$
\begin{array}{c|c|c}
\hbox{Coeff. of} & \hbox {Eq. (\ref{LHS})} & \hbox {Eq. (\ref{RHS})} \\ 
\hline
\un{\tilde\psi}_{N-1}\tilde\phi_{N-1}  &
\begin{array}{c} \gamma^2(N-1) +\\
\sum_{j=1}^{d_1} \alpha_j(N+j-1)\beta_j(N+j-1)
\end{array} &
\begin{array}{c} \gamma^2(N-1) + \\
\sum_{j=1}^{d_2} \alpha_j(N+j-1)\beta_j(N+j-1)
\end{array}  \\
\hline
\un{\tilde\psi}_{N}\tilde\phi_{N} & 
- \gamma^2(N-1) - \sum_{j=1}^{d_1} \alpha_j(N)\beta_j(N)
&
- \gamma^2(N-1) - \sum_{j=1}^{d_2} \alpha_j(N)\beta_j(N)
\\
\hline
\begin{array}{c}
\un{\tilde\psi}_{N+p}{\tilde\phi}_{N+q} \\
p\geq 0 , q\geq 0 \\
p+q \geq 1
\end{array}
& \ds{ -\sum_{l=q+1}^{d_2} \a_l(N+q)\b_{p-q+l} (N+p) } 
& \ds{ -\sum_{l=q+1}^{d_2} \a_l(N+q)\b_{p-q+l} (N+p) } 
\\ 
\hline
\begin{array}{c}
\un{\tilde\psi}_{N-p}\tilde\phi_{N-q} \\
p\geq 1 , q\geq 1 \\
p+q\geq 3
\end{array}
&  \ds{ \sum_{k=0}  \a_{k+p}(N+k) \b_{k+q}(N+k) } 
&  \ds{ \sum_{k=0}  \a_{k+p}(N+k) \b_{k+q}(N+k) } 
\\ 
\hline
\un{\tilde\psi}_{N-1}\tilde\phi_{N} 
& \ds{\gamma(N-1)\a_0(N)-\gamma(N-1)\a_0(N-1)}  
&\begin{array}{c}
\ds{ \sum_{j=1}
\a_{j}(N-1+j)\b_{j-1}(N-1+j)}\\
\ds{-  \sum_{j=1}\a_{j}(N)\b_{j-1}(N-1)}
\end{array}
\\
\hline
\un{\tilde\psi}_N\tilde\phi_{N-1} 
& \begin{array}{c}
\ds{\sum_{j=1} \a_{j-1}(N+j-1) \b_{j}(N+j-1)}\\
\ds {-\sum_{j=1} \a_{j-1} (N-1)\b_{j}(N) }
\end{array}
&\ds { \gamma(N-1) \b_0(N)-\gamma(N-1)\b_0(N-1) }
\\ 
\hline
\begin{array}{c}
\un{\tilde\psi}_{N}{\tilde\phi}_{N- q} \\  q\geq 2
\end{array}
&\begin{array}{c}
\ds{ \sum_{j=0}\!\!
\a_{j}(N+j)\b_{j+q}(N+j)}\\
\ds{ -\sum_{j=-1} \b_{j+q}(N) \a_{j}(N-q)
} \end{array}
&\ds{ -\gamma(N-1)\b_{q-1}(N-1)\ }
\\ 
\hline
\begin{array}{c}
\un{\tilde\psi}_{N+p}\tilde\phi_{N-1} \\
p\geq 1
\end{array}
& \begin{array}{c} 
\ds
{ \sum_{j=-1} \a_j(N+p+j) 
\b_{j+p+1}(N+p+j) }\\
\ds{-\sum_{j=0}
\a_j(N-1)\b_{j+p+1}(N+p) }\end{array}  &
\ds{ \gamma(N-1) \b_p(N+p) }
\\ 
\hline
\begin{array}{c}
\un{\tilde\psi}_{N-1}{\tilde\phi}_{N+ q} \\  q\geq 1
\end{array}
& \gamma(N-1) \alpha_q(N+q)
&\begin{array}{c}
\ds{ \sum_{j=p}
\a_{j}(N+j-1)\b_{j-q-1}(N+j-1)}\\
\ds{ -\sum_{j=0}
\a_{j+q+1}(N+q) \b_{j}(N-1) 
} \end{array}
\\
\hline
\begin{array}{c}
\un{\tilde\psi}_{N-p}\tilde\phi_{N} \\
p\geq 2
\end{array}
& -\gamma(N-1) \a_{p-1}(N-1)
& \begin{array}{c} 
\ds
{ \sum_{j=0} \a_{j+p}(N+j) \b_{j}(N+j) }
\\
\ds{-\sum_{j=-1} \a_{j+p}(N) \b_{j}(N-p) }
\end{array}  
\\ 
\hline
\begin{array}{c}
\un{\tilde\psi}_{N+p}{\tilde\phi}_{N-q} \\
 q\geq 2 ,\ p\geq 0
\end{array}
& 
\begin{array}{c}
\ds
{ \sum_{j=-1} \a_{j} (N+j+p) \b_{j+p+q}(N+j+p)}
\\
\ds{- \sum_{j=-1} \b_{j+p+q}(N+p)\a_{j}(N-q) }
\end{array}
& 0 
\\ 
\hline
\begin{array}{c}
\un{\tilde\psi}_{N-p}\tilde\phi_{N+q} \\
q\geq 0 ,\ p\geq 2 
\end{array}
& 0 &
\begin{array}{c}
\ds{ \sum_{j=q-1} \a_{j+p}(N+j)\b_{j-q}(N+j)}\\
\ds{ - \sum_{j=-1}  \a_{j+p+q}(N+q)\b_{j}(N-p) }
\end{array}
\end{array}
$$
\end{small}
We have defined $\a_j:= 0$ if $j\notin [-1,\dots,d_2]$ and 
$\b_j\equiv 0$ if $j\notin [-1,\dots,d_1]$.
\caption{Comparison of coefficients in eq.(\ref{LHS}) and eq.(\ref{RHS}).}
\label{table}
\end{table}

The second way does not involve any computation; in fact, we can already
conclude that the coefficients of all terms
$\un{\tilde\psi}_{N+p}(y')\tilde\phi_{N+q}(y)$ must agree.  We know that for 
the  polynomial solutions $\{\psi_n(x), \phi_n(y)\}$ and their Fourier-Laplace 
transforms $\{\un\psi_n(y'), \un\phi_n(x')\}$ the two expressions are the same 
because
\bea
&&\le(\pa_{y'}+\pa_{y}\ri) 
 \le(\m{\un\Psi}^{N\!-\!1}(y'),\Bmat^N \m{\Phi}_N (y)\ri)
 =
\le(\pa_{y'}+\pa_y\ri) (y'-y) \sum_{n=0}^{N-1} \un\psi_n(y')\phi_n(y)
 \cr
&&= (y'-y)\le(\pa_{y'}+\pa_y\ri) \sum_{n=0}^{N-1}
\un\psi_n(y')\phi_n(y)=\le(y' -y\ri) 
 \le(\m{ \un\Psi}_N (y'), (\Amat^N)^t \m{\Phi}^{N\!-\!1}(y)\ri)\ ,
\eea
where we have used the generalized Christoffel--Darboux formulae  and the
identity
\be
\le[(y'-y),\le(\pa_{y'}+\pa_y\ri) \ri]=0 \ .
\ee
Since the $\un\psi_n(y')$'s are linearly independent  and so are the
$\phi_n(y)$'s, the functions 
\be
F_{pq}(y',y) :=
\un\psi_{N\!+\!p} (y')\phi_{N\!+\!q}(y)
\ee
  are also linearly
independent.
Considering eqs. (\ref{LHS}) and (\ref{RHS}) as linear equalities for the 
$F_{pq}(y',y)$'s, one concludes that the coefficient in the two equations must 
be equal. \hfill Q.E.D.

Before stating the next result, we define new pairings by studying the
effect of deformations on the kernels. By using the same
Christoffel--Darboux trick one easily computes
\bea
&& \frac {\pa}{\pa u_K} \sum_{n=0}^{N-1} \psi_n(x)\un\phi_n(x') =
\sum_{j=1}^K \sum_{l=1}^j U^{K}_j(N\!-\!l) \psi_{N\!-l+\!j}
\un\phi_{N\!-l}  \ , \\
&& \frac {\pa}{\pa v_K} \sum_{n=0}^{N-1} \psi_n(x)\un\phi_n(x') =
\sum_{j=1}^K \sum_{l=1}^j V^{K}_j(N\!-\!l) \un\phi_{N\!-l+\!j}
\psi_{N\!-l}  \ .
\label{deformpairings}
\eea
This pairing does not change if we take
Fourier--Laplace  transforms of  $\Psi$ or $\Phi$, so we can easily write
  the deformations of the two kernels $K_{11}$
 and $K_{22}$.
\bp
\label{mainlemma2}
For any two sequences satisfying both the deformation equations and
the $x$-recursion relations we have
\bea
&& \frac{\pa}{\pa u_K}
\le(\m{\un{\tilde \Psi}}^{N\!-\!1} (y'),
\Bmat^N\m{\tilde\Phi}_{N} (y)\ri) = (y'-y)
\le( \sum_{j=1}^K \sum_{l=1}^j U^{K}_j(N\!-\!l) \un{\tilde\psi}_{N\!-l+\!j}
{\tilde\phi}_{N\!-l}\ri)   \\  
&& \frac{\pa}{\pa v_J}
\le(\m{\un{\tilde\Psi}}^{N\!-\!1} (y'),
\Bmat^N\m{\tilde \Phi}_{N} (y)\ri) = 
 (y'-y)\le(\sum_{j=1}^J \sum_{l=1}^j V^{J}_j(N\!-\!l) {\tilde\phi}_{N\!-l+\!j}
\un{\tilde \psi}_{N\!-l} \ri) \\
&& \frac{\pa}{\pa u_K}
\le(\m{\un\Phi}^{N\!-\!1} (x'),
\Amat^N\m{\Psi}_{N} (x)\ri) = 
(x'-x)  \le(\sum_{j=1}^K \sum_{l=1}^j U^{K}_j(N\!-\!l)\tilde  \psi_{N\!-l+\!j}
\un{\tilde\phi}_{N\!-l} \ri)\  \\  
&& \frac{\pa}{\pa v_J}
\le(\m{\un\Phi}^{N\!-\!1} (x'),
\Amat^N \m{\Psi}_{N} (x)\ri) = (x'-x)
\le( \sum_{j=1}^J \sum_{l=1}^j V^{J}_j(N\!-\!l) \un{\tilde\phi}_{N\!-l+\!j}
\tilde\psi_{N\!-l} \ri) \ .
\eea
\ep
{\bf Proof}. We prove only one identity, the others being completely similar.  
If the two sequences consist of the orthogonal quasi-polynomials (and the
corresponding Fourier-Laplace transforms), then the equality follows
immediately from: 
\bea
\frac{\pa}{\pa u_K}
\le(\m{\un\Psi}_{N\!-\!1} (y'),
\Bmat^N \m{\Phi}^{N}(y)\ri) &=&\frac{\pa}{\pa u_K}
(y'-y)\sum_{n=0}^{N-1} \un\psi_n(y')\phi_n(y) \cr
&=&   
(y'-y)\le(
\sum_{j=1}^K \sum_{l=1}^j U^{K}_j(N\!-\!l) \un\psi_{N\!-l+\!j}
\phi_{N\!-l} \ri) \ .
\eea
Expanding both sides by means of the recursion relations and the
deformation equations in order to obtain linear expressions in
$F_{p,q}(y',y)$ in the LHS and RHS one concludes that
the coefficients must be the same. But the same final expression
relies only on the recursion relations, and hence the equality holds
for any pair of sequences of functions $\un{\tilde\psi}_n(y')$ and $
\tilde\phi_n(y)$ satisfying these same recursion relations (by the same 
argument as in the proof of Prop. \ref{mainlemma}).
 \hfill  Q.E.D.

\bt
\label{maintheorem}
If  $\{\un{\tilde\psi}_n(y)\}_{n\in \N}$ and $\{\tilde\phi_n(y)\}_{n\in\N}$ (or
$\{\un{\tilde\phi}_n(x)\}_{n\in \N}$ and $\{\tilde\psi_n(x)\}_{n\in\N}$)  are
 arbitrary pairs of sequences of functions satisfying the recursion
relations (\ref{recrelbar1}), (\ref{recrel3}), (resp. (\ref{recrelbar3})), 
(\ref{recrel1}), the differential relations (\ref{recrelbar2}), 
(\ref{recrel4}), (resp. (\ref{recrelbar4}), (\ref{recrel2}))  and the 
deformation equations (\ref{defpsinuK})-(\ref{defphinvJ}), then the bilinear 
expressions
\bea
&& \tilde f_N(y):= 
\le(\m{\un{\tilde\Psi}}^{N\!-\!1} (y), \Bmat^N \m{\tilde\Phi}_N(y)\ri)\ , \\
&& \tilde g_N(x):=
\le(\m{\un{\tilde\Phi}}^{N\!-\!1} (x), \Amat^N \m{\tilde\Psi}_N(x)\ri) \ ,
\eea
are independent of  $y$ (resp. $x$)  and $N$, and also are constant in
the deformation parameters $\{u_K, v_J\}$.
\et
{\bf Proof}. 
Using  Prop. \ref{mainlemma} and setting $y=y'$ we find at once that
$$
\frac d{dy} \tilde f_N(y)= 0,
$$ 
i.e. $f_N(y)=f_N$ does not depend on $y$. A similar computation shows that
$\tilde g_N(x)$ is independent of $x$.  Now we also compute (for, say, $M<N$)
\be
\le(\m{\un{\tilde\Psi}}^{N\!-\!1}
(y), \Bmat^N \m{\tilde\Phi}_N (y')\ri) -
\le(\m{\un{\tilde\Psi}}^{M\!-\!1} (y),\Bmat^M \m{\tilde\Phi}_M  (y')\ri) =
(y-y')\sum_{n=M}^{N-1} \un{\tilde\psi}_n(y')\tilde\phi_n(y)\ .
\ee 
Letting $y=y'$ we obtain $\tilde f_N= \tilde f_M$, and similarly for
$\tilde g_N$
To prove the independence of the deformation parameters $u_K$ and $ v_J$, we 
use Prop. \ref{mainlemma2} in a similar way to the above and
then again set $y=y'$ (or $x=x'$) to conclude the proof.  \hfill Q.E.D.

Theorem \ref{maintheorem} allows us to conclude that we can
choose fundamental systems of solutions to the pairs of dual
differential-difference-deformation equations normalized
in such a way that the pairing gives the identity matrix.
\bc
\label{maincorollary}
There exist two pairs of sequences of fundamental matrix solutions to
the difference--differential--deformation equations 
(\ref{shiftpsi})-(\ref{dxphi_bar}), (\ref{defUVPsi})-(\ref{defUVPhi_bar})
$\ds { (\m{\bf\Psi}_N,\m{\bf\un\Phi}^{N-1}) }$, $\ds
{ (\m{\bf\Phi}_N,\m{\bf\un\Psi}^{N-1}) }$ such that 
\be
\left(\m{\bf\un\Phi}^{N-1} ,  \Amat^N  \m{\bf\Psi}_N)\right)\equiv \1\ ,\qquad 
\left(\m{\bf\un\Psi}^{N-1} ,  \Bmat^N  \m{\bf\Phi}_N\right) \equiv \1 \ .
\ee
\ec

We conclude with the following theorem.
\bt 
\label{system_duality}
The differential-deformation systems 
\bea
&& 
\pa_y \m{\bf \Phi}_N = - \m{D_2}^N(y)\m{\bf \Phi}_N ,  \quad
 \pa_{u_K}\m{\bf \Phi}_N =- {\m{\bf U}^{N,\Phi}}_{\!\!K} \m{\bf \Phi}_N , 
\quad
\pa_{v_J}\m{\bf \Phi}_N = {\m{\bf V}^{N,\Phi}}_{\!\!J}\m{\bf \Phi}_N , \\
&&
\pa_y \m{\bf\un \Psi}^{N\!-\!1} =
\m{\bf\un \Psi}^{N\!-\!1} \m{{\un D}_2}^N(y)  , \quad
\pa_{u_K}\m{\bf\un \Psi}^{N\!-\!1} =
\m{\bf\un \Psi}^{N\!-\!1}{\m{\un{\bf U}}^{N,\Psi}}_{\!\!K} , 
\quad
\pa_{v_J}\m{\bf\un \Psi}^{N\!-\!1} =
- \m{\bf\un \Psi}^{N\!-\!1} {\m{\un{\bf V}}^{N,\Psi}}_{\!\!J} ,
\eea
for $ K=1 \dots d_1\!+\!1;\ J=1 \dots d_2\!+\!1,$
 are put in duality by the matrix $\ds{\Bmat^N}$. 
\bea
&&
\Bmat^N \m{D_2}^N(y) = \m{\un D_2}^N(y)\Bmat^N   \ ,\\
&& \pa_{u_K} \Bmat^N =  \Bmat^N  {\m{\bf
U}^{N,\Phi}}_{\!\!K}(y) -  {\m{\un{\bf U}}^{N,\Psi}}_{\!\!K}(y)\Bmat^N  \ ,\\
&&\pa_{v_J} \Bmat^N =  {\m{\un{\bf V}}^{N,\Psi}}_{\!\!J}(y)\Bmat^N  -
\Bmat^N  {\m{\bf V}^{N,\Phi}}_{\!\!J}(y)\ .
\eea
In particular, since the matrices $\ds{{\m{\un D}^N}_2(y)}$ and 
$\ds{ {\m{ D}^N}_2(y)}$ are conjugate to each other, their spectral curves
are the same.
Similarly, the differential-deformation systems
\bea
&&
\pa_x \m{\bf \Psi}_N = - \m{D_1}^N(x)\m{\bf \Psi}_N ,  \quad 
 \pa_{u_K}\m{\bf \Psi}_N = {\m{\bf U}^{N,\Psi}}_{\!\!K} \m{\bf \Psi}_N , \quad
\pa_{v_J}\m{\bf \Psi}_N = - {\m{\bf V}^{N,\Psi}}_{\!\!J}\m{\bf \Psi}_N  \ ,
\\ &&
\pa_x \m{\bf\un \Phi}^{N\!-\!1} = 
\m{\bf\un \Phi}^{N\!-\!1} \m{{\un D}_1}^N(x)  , \quad 
\pa_{u_K}\m{\bf\un \Phi}^{N\!-\!1} =
- \m{\bf\un \Phi}^{N\!-\!1}{\m{\un{\bf U}}^{N,\Phi}}_{\!\!K} , \quad    
\pa_{v_J}\m{\bf\un \Phi}^{N\!-\!1} = 
\m{\bf\un \Phi}^{N\!-\!1} {\m{\un{\bf V}}^{N,\Phi}}_{\!\!J}  \ , 
\eea
$K=1 \dots d_1\!+\!1;\ J=1 \dots d_2\!+\!1,$ are put in duality by the matrix
$\ds{\Amat^N}$,
\bea
&&
\Amat^N \m{D_1}^N(x) = \m{\un D_1}^N(x)\Amat^N  \ , \\
&& \pa_{v_J} \Amat^N =  \Amat^N  {\m{\bf
V}^{N,\Psi}}_{\!\!J}(x) -  {\m{\un{\bf V}}^{N,\Phi}}_{\!\!J}(x)\Amat^N  \ ,\\
&&\pa_{u_K} \Amat^N =  {\m{\un{\bf U}}^{N,\Phi}}_{\!\!K}(x)\Amat^N  -
\Amat^N  {\m{\bf U }^{N,\Psi}}_{\!\!K}(x)\ ,
\eea
and hence the spectral curves of $\ds{{\m{\un D}^N}_1(x)}$ and 
$\ds{ {\m{ D}^N}_1(x)}$ are also the same. 
\label{dualitythm}
\et
{\bf Proof}.
The three relations follow easily by taking a fundamental system of
solutions for the two compatible  differential--deformation systems and
using Theorem \ref{maintheorem}.  \hfill Q.E.D.

This theorem together with Prop. \ref{accordion} proves that the four
spectral curves
\bea
\begin{array}{ccc}
 \ds{
\det\le[ y\1- {\m{\un D}^N}_2(x)\ri]=0}
& \ds{\m{\longleftrightarrow}^{\rm Prop. \ \ref{accordion}}} 
&\ds{\det\le[ x\1- {\m{D}^N}_1(y)\ri]=0 } \\[5pt]
\updownarrow {\rm Thm.\ \ref{maintheorem}} & &\updownarrow {\rm Thm.\
\ref{maintheorem}} \\[5pt]
\ds{\det\le[ y\1- {\m{D}^N}_2(x)\ri]=0} &
\ds{\m{\longleftrightarrow}^{\rm Prop. \ \ref{accordion}}}  &
\ds{ \det\le[ x\1- {\m{\un D}^N}_1(y)\ri]=0}
\end{array}
\eea
all coincide.
\bigskip

\subsection{Concluding remarks}

  In this work, the main results concern the compatibility of the difference-
differential-deformation systems arising from the ``folding'' procedure 
(Proposition \ref{shift_dx_compatibility}), and the resulting spectral duality 
Theorems \ref{maintheorem} and \ref{system_duality}. The constancy of the
bilinear pairings between solutions given by Corollary \ref{maincorollary} may
be viewed as a form of the bilinear relations for Baker functions which imply
the Hirota bilinear equations for the associated tau function of eq.
(\ref{deftau}).

  Another consequence of this compatibility is the fact that the (generalized) 
monodromy of the covariant derivative operators 
${d \over d x} - \ds{\m{D_1}^N}(x)$, and ${d \over d y} - \ds{\m{D_2}^N}(y)$ is 
independent of both the continuous deformation parameters $\{u_K, v_J\}$ and 
the  integer $N$; i.e., we have a dual pair of  differential operators families
whose  coefficients satisfy differential equations in the parameters $\{u_K,
v_J\}$ and difference equations in the discrete parameters $N$ that 
generate isomonodromic deformations.  Associated to such isomonodromic
deformation equations, there is a sequence of isomonodromic tau functions in 
the sense of refs \cite{JMU, JM}. However, since the highest terms of the
polynomial matrices $\ds{\m{D_1}^N}(x)$ and $\ds{\m{D_2}^N}(y)$  have a very
degenerate spectrum (in fact, they have rank 1), the standard definition of
the isomonodromic tau function does not apply. To introduce a suitable 
definition for this situation, an analysis of the formal asymptotics
at $x=\infty$ (or $y=\infty$) is required.  Also, the systems of
Proposition \ref{shift_dx_compatibility}  represent in a sense, the ``vacuum'' 
isomonodromic deformation systems associated with the Fredholm  kernels
appearing in Proposition \ref{generalDC}. When the corresponding integral
operator is supported on a union of intervals, the computation of its resolvent
is equivalent to a Riemann-Hilbert problem with discontinuities given across
these cuts \cite{IIKS}. The resulting ``dressed'' Baker  functions determine
isomonodromic families of covariant derivative operators having, in addition to
polynomial parts, poles at the endpoints of the intervals, which may be
viewed as new deformation parameters. The associated isomonodromic tau functions
are given by the Fredholm determinants of the integral operator supported on
the union of intervals.
\cite{HI,IIKS}.  

  The study of the formal asymptotics associated to the vacuum systems, the 
corresponding isomonodromic tau functions and the relation of these to the 
spectral invariants will be developed in a subsequent work (\cite{BEH2}), 
as will the study of the $N\rightarrow \infty$ asymptotics of the biorthogonal 
polynomials and associated Fredholm kernels.

\bigskip
\bigskip

\appendix
\section{Appendix: Multi-matrix model}

The ``multi-matrix-model'' is a generalization of the 2-matrix-model, which 
was introduced in the context of string theory and conformal field theory
\cite{DouglasKdV, ZJDFG}, and has been extensively studied 
\cite{Mehta, Mehtaorthpol, ZJDFG}.  Our notations in the following
mainly follow \cite{eynardchain}.  Calculations of spectral statistics
in this model again involves biorthogonal polynomials which obey 
linear differential systems of finite rank. It will be shown in this
appendix that these also satisfy an extended form of the spectral duality 
relations derived for the 2-matrix case. The results will be summarized, but 
only a brief sketch of the proofs will be indicated.

\bigskip 

Consider $m\geq 2$ random hermitian $N\times N$ matrices $M_1, M_2, \dots ,
M_m$, with the measure

\be
{\rm d}\mu = \prod_{k=1}^m {\rm e}^{-{\rm Tr}\, V_k(M_k)}\,\,
\prod_{k=1}^{m-1} {\rm e}^{{\rm Tr}\, M_k M_{k+1}} \prod_{k=1}^M {\rm d}\, M_k
\ ,
\ee
where ${\rm d}M_k$ is the standard Lebesgue measure for hermitian matrices, and
the potentials $V_k$, $k=1\dots m$, are polynomials of degrees $d_k+1$, with
coefficients
 \be
V_k(x) = u_{k,0} + \sum_{l=1}^{d_k+1}\frac {u_{k,l}}{l} x^l \ .
\ee

\medskip

As in  the 2-matrix case, all the correlation functions and statistical 
properties of the eigenvalues of the $m$ matrices can be expressed  in terms 
of determinants involving $m^2$ Fredholm integral kernels, which are 
constructed from an infinite sequence  of biorthogonal polynomials 
and their integral transforms \cite{eynardmehta}. In this case, the
biorthogonal polynomials
\be
\pi_n(x) = x^n + \dots \qquad , \qquad
\sigma_n(y) = y^n+\dots
\ee
are defined to be orthogonal in the following sense:
\be
\int\int\dots \int \,\, {\rm d}x_1\,{\rm d}x_2\,\dots\,{\rm d}x_{m-1}\,
{\rm d}x_m \,\,    \prod_{k=1}^m {\rm e}^{- V_k(x_k)}\,\, \prod_{k=1}^{m-1}
{\rm e}^{ x_k x_{k+1}} \,\,\pi_n(x_1)   \,\,\sigma_l(x_m)\, = h_n \delta_{nl} 
\ ,
\ee
where the integral is convergent on the real axis if all the degrees $d_k+1$ 
are even, and the leading coefficients are positive. Otherwise we need to 
consider other integration paths in the complex plane, without boundaries, so
that integration by parts may be done. This uniquely determines the
polynomials
$\pi_n$ and $\sigma_n$ for all $n$.

\medskip

From $\pi_n$, we define the following $m$ sequences of functions
$\le\{ \psi_{1,n}\ri\}_{n=0\dots \infty} ,
\dots \le\{ \psi_{m,n}\ri\}_{n=0\dots \infty}$ 
\bea 
\psi_{1,n}(x) & := & {1\over \sqrt{h_n}} \pi_n(x) {\rm e}^{-V_1(x)}  \ ,
\cr
 \psi_{2,n}(x) & := & \int {\rm d}y \
\psi_{1,n}(y) \  {\rm e}^{x y}  \ ,  \cr
 \psi_{k+1,n}(x) & := & \int {\rm d}y \ 
\psi_{k,n}(y) \ {\rm e}^{x y}\  {\rm e}^{-V_{k}(y)} \quad
{\rm for}\quad, m-1\geq k\geq 2 \ , \label{Adefpsi}
\eea
and from $\sigma_n$, the following $m$ sequences of functions
$\le\{\phi_{1,n}\ri\}_{n=0\dots \infty},\dots 
\le\{\phi_{m,n}\ri\}_{n=0\dots \infty}$
\bea
\phi_{m,n}(x) &:=& {1\over \sqrt{h_n}} \sigma_n(x) {\rm e}^{-V_m(x)}
\cr
 \phi_{k-1,n}(x) &:=& \int {\rm d}y\ 
\phi_{k,n}(y) \ {\rm e}^{x y}\,\, {\rm e}^{-V_{k-1}(x)}
\qquad {\rm for}\quad m\geq k\geq 3 \cr
 \phi_{1,n}(x) &:=& \int {\rm d}y\ 
\phi_{2,n}(x) \ {\rm e}^{x y}  \ , \label{Adefphi}
\eea
which are dual bases for the respective spaces they span:
\be\label{Apairing}
\int {\rm d}x\,\, \psi_{k,n}(x) \,\phi_{k,l}(x) 
=\delta_{nl} \ .
\ee

\subsection{ Recursion relations}

We define the semi-infinite matrices $P_k$ and $Q_k$ for each $k=1,\dots ,m$,
such that \cite{twomatrixHeisenberg, eynardchain} 
\be
x \psi_{k,n}(x) = \sum_{l} {Q_k}_{(n,l)} \psi_{k,l} (x)
\ , \qquad
{\partial \over \partial x} \psi_{k,n}(x) = \sum_{l} {P_k}_{(n,l)}
 \psi_{k,l} (x) \ ,
\ee
where it will be shown below that these only involve finite sums.
From the pairing (\ref{Apairing}), and integration by parts, we have
\be
x_k \phi_{k,n}(x) = \sum_{l} {Q^t_k}_{(n,l)} \phi_{k,l} (x)
\ , \qquad
{\partial \over \partial x} \phi_{k,n}(x) = - \sum_{l} {P^t_k}_{(n,l)}
 \phi_{k,l} (x)  \ .
\ee
Note that these matrices all satisfy Heisenberg relations
\cite{twomatrixHeisenberg} 
\be
[P_k,Q_k] = {\rm \1}.
\ee
Using the definitions of $\psi_{k,n}$ and $\phi_{k,n}$, we find the following
relationships between them.
\be
P_{k+1} = Q_k \quad  k\in [1,m-1]
\ , \qquad
-P_{k} = Q_{k+1} - V'_{k}(Q_{k}) \quad  k\in [2,m-1] \ ,
\qquad  -P_{1} = Q_2 \ .
\ee
In particular, this implies that
\be\label{ArelQ}
Q_{k-1}+Q_{k+1} = V'_k(Q_k) \qquad {\rm for}\,\,\, k\in [2,m-1] \ .
\ee
These relations are enough to ensure that all the matrices $Q_k$ and $P_k$
are of finite band type.

\bp{The matrix $Q_k$ has $r_k$ bands above the principal diagonal and $s_k$
bands below the principal diagonal, with}
\bea
r_1=1 & {\rm and} & r_k = \prod_{l=1}^{k-1} d_l  \quad {\rm for}
\  k\in [2,m] \ , \\
s_m=1 & {\rm and} & s_k = \prod_{l=k+1}^{m} d_l  \quad {\rm for}
\  k\in [1,m-1]  \ .
\eea

\noindent
{\bf Proof:}
$Q_1$ multiplies the vector of polynomials $\{\pi_{n}(x)\}_{n=0\dots \infty}$ by
$x$, and can therefore raise the degree at most by $1$; i.e. $Q_1$
has at most one line above diagonal.
From the same argument, since the multiplication by $P_1+V'_1(Q_1)$ takes the
derivative of the vector polynomial $[\pi_{n}(x)]_{n=0\dots \infty}$ with
respect to $x$, it must lower the degree, therefore $P_1$ has at most
$d_1=\deg V'_1$ lines above diagonal and so does $Q_2=-P_1$. Using
\ref{ArelQ} recursively, it follows that $Q_k$ has at most $r_k$ lines above
diagonal. Repeating the argument with the polynomials $\{\sigma_n\}$, we see 
that $Q^t_k$ has at most $s_k$ lines above the diagonal. \hfill  Q.E.D.
\ep

Denoting
\be
\alpha_{k,l}(n) := {Q_k}_{(n,n+l)} \ ,
\ee
 the recursion relations may be written componentwise as 
\be
x \psi_{k,n}(x) = \sum_{l=-s_k}^{+r_k} \alpha_{k,l}(n) \psi_{k,n+l}(x)
\ , \qquad
x \phi_{k,n}(x) = \sum_{l=-s_k}^{+r_k} \alpha_{k,l}(n-l) \phi_{k,n-l}(x) \ ,
\ee
which implies
\be
{\partial \over \partial x} \psi_{1,n}(x)   = 
- \sum_{l=-s_{2}}^{+r_{2}} \alpha_{2,l}(n) \psi_{1,n+l}(x)  \ , \quad
{\partial \over \partial x} \psi_{k,n}(x)   = 
\sum_{l=-s_{k-1}}^{+r_{k-1}} \alpha_{k-1,l}(n) \psi_{k,n+l}(x) \ , 
\ k\in [2,m] \ , 
\ee
and
\be
{\partial \over \partial x} \phi_{1,n}(x)   = 
 \sum_{l=-s_{2}}^{+r_{2}} \alpha_{2,l}(n-l) \phi_{1,n-l}(x)  \ , \quad
{\partial \over \partial x} \phi_{k,n}(x)   = 
- \sum_{l=-s_{k-1}}^{+r_{k-1}} \alpha_{k-1,l}(n-l) \phi_{k,n-l}(x)  \ , 
k\in [2,m] \ .
\ee

Note that
\be
\alpha_{1,1}(n) = \alpha_{m,-1}(n+1) = \gamma(n) = \sqrt{h_{n+1}\over h_n} \ .
\ee

\subsection{Folding}

Again, it is possible to ``fold'' these recursion relations to form finite
rank linear differential systems with polynomial coefficients.
For each $k$,  define the following windows of size $r_k+s_k$.
\be
\mathop{\Psi_k}_{N} = \pmatrix{\psi_{k,N-s_k} \cr \vdots \cr \psi_{k,N+r_k-1}}
\ , \qquad
\mathop{\Phi_k}_{N} = \pmatrix{\phi_{k,N-r_k} & \dots & \phi_{k,N+s_k-1}} \ .
\ee

\noindent
{\bf Shift operators:}
For each $k$,  define ``ladder'' matrices of size $r_k+s_k$,
\bea
&&\mathop{a_k}_N (x) =
\pmatrix{
0 & 1 & 0 & 0 \cr
0 & 0 & \ddots & 0 \cr
0 & 0 & 0 & 1 \cr
-{\alpha_{k,-s_k}(N)\over \alpha_{k,r_k}(N)} & \dots & {x-\alpha_{k,0}(N)\over
\alpha_{k,r_k}(N)} \dots & -{\alpha_{k,r_{k}-1}(N)\over \alpha_{k,r_k}(N)}
}
\cr
&&\mathop{\tilde{a}_k}_N (x) =
\pmatrix{
-{\alpha_{k,r_{k}-1}(N-r_{k}+1)\over \alpha_{k,r_k}(N-r_k)} & 1 & 0 & 0 \cr
{x-\alpha_{k,0}(N)\over \alpha_{k,r_k}(N-r_k)} & 0 & \ddots & 0 \cr
\vdots & 0 & 0 & 1 \cr
-{\alpha_{k,-s_k}(N+s_k)\over \alpha_{k,r_k}(N-r_k)} & 0 & 0 & 0
}
\eea
which implement the shifts\footnote{In the $m=2$ case, we had ${\bf a}=a_1$,
$\underline{\bf a}=\tilde{a}_1$, ${\bf b}=(\tilde{a}_2^t)^{-1}$,
$\underline{\bf b}=(a_2^t)^{-1}$.} in $N$. 
\be
\mathop{a_k}_N (x)
\mathop{\Psi_k}_{N} = \mathop{\Psi_k}_{N+1} \ , \qquad
\mathop{\Phi_k}_{N} \mathop{\tilde{a}_k}_N (x) = \mathop{\Phi_k}_{N-1} \ .
\ee

\noindent
{\bf Differential systems:}
The differential systems\footnote{Notice that the notations are
changed for $m=2$. $D_2$ is now what we called $-\underline{D}_2^t$ and
$\tilde{D}_2$ is what we previously called $-D_2^t$.} satisfied by the vectors
$\mathop{\Psi_k}_{N}(x)$ and $\mathop{\Phi_k}_{N}(x)$ are:
\be
{\ds{\partial\over \partial x}
 \mathop{\Psi_k}^{N}(x) = \mathop{D_k}^N(x) \mathop{\Psi_k}^{N}(x)
\ , \qquad
{\partial\over \partial x} \mathop{\Phi_k}_{N}(x) =
 - \mathop{\Phi_k}_{N}(x) \mathop{{\tilde{D}}_k}^N(x)}
\label {MMdx}
\ee
where ${\ds \mathop{D_k}^N(x)} $ and ${\ds \mathop{{\tilde{D}}_k}^N(x)}$ are
matrices of size
$r_k+s_k$, and with polynomial coefficients of degree at most $d_k$ in $x$.

The matrices ${\ds \mathop{D_k}^N(x)} $ are given by:
\bea
 &&\mathop{D_1}^N(x) =  - \ds{{\mathop\alpha^{1 N}}_{2,0}} - \sum_{
l=1}^{r_{2}}
{\mathop\alpha^{1 N}}_{2,l} \prod_{j=1}^{l} \mathop{a_1}_{N+l-j} (x)
- \sum_{l=1}^{s_{2}} \ds{{\mathop\alpha^{1 N}}_{2,-l}}
\prod_{j=0}^{l-1} \left(\mathop{a_1}_{N-l+j} (x)\right)^{-1} \cr
 &&\mathop{D_k}^N(x) =  \ds{{\mathop\alpha^{k N}}_{k-1,0}}+
\sum_{l=1}^{r_{k-1}} {\mathop\alpha^{k N}}_{k-1,l} \prod_{j=1}^{l}
\mathop{a_k}_{N+l-j} (x) + \sum_{l=1}^{s_{k-1}} {\mathop\alpha^{k N}}_{k-1,-l}
\prod_{j=0}^{l-1} \left(\mathop{a_k}_{N-l+j} (x)\right)^{-1} \cr
&&{\rm for} \  m\geq k\geq 2 \ ,
\eea
where 
\be
{\mathop\alpha^{k N}}_{j,l} := {\rm diag}\, (\alpha_{j,l}(N-s_k) , 
\dots , \alpha_{j,l}(N+r_k-1)) \ .
\ee

Similarly:
\bea
\mathop{\tilde{D}_1}^N(x) &&=  - \ds{\mathop{\tilde\alpha}^{1 N}}_
{2,0} \,\, -
\sum_{l=1}^{r_{k-1}} \left(\prod_{j=0}^{l-1} \mathop{\tilde{a}_1}_{N-j}
(x) \right) \,\ds{\mathop{\tilde\alpha}^{1 N}}_{2,l}  \,\,
 - \sum_{l=1}^{s_{k-1}}
\left( \prod_{j=1}^{l} \left(\mathop{\tilde{a}_1}_{N+j} (x)\right)^{-1} \right)
\, \mathop{\tilde\alpha}^{1 N}_{2,-l} \cr
 \mathop{\tilde{D}_k}^N(x) &&=  \ds{\mathop{\tilde\alpha}^{k N}}_{k-1,0} \,\,
 + \sum_{l=1}^{r_{k-1}} \left(\prod_{j=0}^{l-1} \mathop{\tilde{a}_k}_{N-j}
(x) \right) \,\ds{\mathop{\tilde\alpha}^{k N}}_{k-1,l}  \,\, 
+ \sum_{l=1}^{s_{k-1}}  \left( \prod_{j=1}^{l} \left(\mathop{\tilde{a}_k}_{N+j}
(x)\right)^{-1} \right)
\, \mathop{\tilde\alpha}^{k N}_{k-1,-l} \cr
{\rm for} && m\geq k\geq 2
\eea
where 
\be
\ds{\mathop{\tilde\alpha}^{k N}}_{j,l} : ={\rm diag}\, (\alpha_{j,l}(N-l-r_k) ,
 \dots , \alpha_{j,l}(N-l+s_k-1)) \ .
\ee

\bigskip

{\bf Christoffel--Darboux matrices:}

Consider the kernel
\be
{\mathop{K}_N}_{k,k} (x,y) = \sum_{n=0}^{N-1} \phi_{k,n}(y) \psi_{k,n}(x) \ .
\ee

The generalization of the Christoffel--Darboux theorem for this kernel reads
\bea
(x-y) {\mathop{K}_N}_{k,k}(x,y) 
&=& \sum_{l=1}^{r_k} \sum_{j=1}^l \alpha_{k,l}(N-j) \psi_{k,N-j+l} \phi_{k,N-j}
- \sum_{l=1}^{s_k} \sum_{j=1}^l \alpha_{k,-l}(N-j+l) \psi_{k,N-j}
\phi_{k,N-j+l} \cr
&=& \mathop{\Phi_k}_{N}(x) \mathop{ \mathbb A_k}_N  \mathop{\Psi_k}_{N}(y) \ .
\eea
where ${\ds \mathop{\mathbb A_k}_N}$ is the $(r_k+s_k)\times (r_k+s_k)$ matrix:
\be
\mathop{\mathbb A_k}_N = \pmatrix{
0 & \dots & 0& \alpha_{k,r_k}(N-r_k) & 0& 0\cr
\vdots & \dots &  \vdots & \vdots& \vdots& \vdots\cr
0 & \dots & 0& \vdots & \ddots & 0\cr
0 & \dots & 0& \alpha_{k,1}(N-1) & \dots & \alpha_{k,r_k}(N-1) \cr
-\alpha_{k,-s_k}(N) & \dots & -\alpha_{k,-1}(N) & 0& \dots& 0\cr
0 & \ddots & \vdots & 0 & \dots& 0\cr
\vdots & \vdots &  \vdots & \vdots& \vdots& \vdots\cr
0 & 0& -\alpha_{k,+s_k-1}(N+s_k-1)& 0& 0& 0\cr
}
\ee

There is also a ``differential Christoffel--Darboux theorem'', for $k>1$:
\bea
(\partial_x + \partial_y) {\mathop{K}_N}_{k,k}(x,y) 
=&& \sum_{l=1}^{r_{k-1}} \sum_{j=1}^l \alpha_{k-1,l}(N-j) \psi_{k,N-j+l}
\phi_{k,N-j} \cr
&& - \sum_{l=1}^{s_{k-1}} \sum_{j=1}^l \alpha_{k-1,-l}(N-j+l)
\psi_{k,N-j} \phi_{k,N-j+l} \cr
=&& \mathop{\hat\Phi_k}_{N}(x) \mathop{\mathbb A_{k-1}}_N 
\mathop{\hat\Psi_k}_{N}(y)
\ ,
\eea
where
\be
\mathop{\hat\Psi_k}_{N} = \pmatrix{\psi_{k,N-s_{k-1}} & \dots &
\psi_{k,N+r_{k-1}-1}}^t \ , \qquad
\mathop{\hat\Phi_k}_{N} = \pmatrix{\phi_{k,N-r_{k-1}} & \dots &
\phi_{k,N+s_{k-1}-1}} \ ,
\ee
and for $k=1$,
\bea
(\partial_x + \partial_y) {\mathop{K}_N}_{1,1}(x,y) 
&=& -\sum_{l=1}^{r_{2}} \sum_{j=1}^l \alpha_{2,l}(N-j) \psi_{1,N-j+l}
\phi_{1,N-j} \cr
&& \  + \sum_{l=1}^{s_{2}} \sum_{j=1}^l \alpha_{2,-l}(N-j+l)
\psi_{1,N-j} \phi_{1,N-j+l} \cr
&=& -\mathop{\hat\Phi_1}_{N}(x) \mathop{\mathbb A_{2}}_N 
\mathop{\hat\Psi_1}_{N}(y) \ ,
\eea
where
\be
\mathop{\hat\Psi_1}_{N} = \pmatrix{\psi_{1,N-s_{2}} & \dots &
\psi_{1,N+r_{2}-1}}^t \ , \qquad
\mathop{\hat\Phi_1}_{N} = \pmatrix{\phi_{1,N-r_{2}} & \dots &
\phi_{1,N+s_{2}-1}} \ .
\ee

\subsection{Duality}

All the systems $\ds{\mathop{D_k}^N(x)}$ and
$\ds{\mathop{{\underline{D}}_k}^N(x)}$ have the same spectral curve.
This result follows again in two steps.

$\bullet$
For each $k$, we have:
\be\label{dualconjug}
\mathop{{\tilde{D}}_k}^N(x) \, \mathop{\mathbb A_k}_N = \mathop{\mathbb A_k}_N
\, \mathop{D_k}^N(x)
\ee
which implies that 
\be
\det{\left( y {\rm \1} - \mathop{D_k}^N(x)  \right)}
= \det{\left( y {\rm \1} - \mathop{{\tilde{D}}_k}^N(x)  \right)} \ .
\ee

$\bullet$ 
The relationship between spectral curves for different $k$ is:
\be\label{dualtensor}
\det{\left( x {\rm \1} -  \mathop{D_{k+1}}^N(y) \right)}
\propto \det{\left( y {\rm \1} -V'_k(x) {\rm \1} +  \mathop{D_k}^N(x)
\right)} \qquad {\rm for} \,\, k>1  \ ,
\ee
and
\be
\det{\left( x {\rm \1} -  \mathop{D_2}^N(y) \right)}
\propto \det{\left( y {\rm \1} +  \mathop{D_{1}}^N(x) \right)} \ .
\ee

\bigskip

\noindent
{\bf Proof:}
 We prove \ref{dualconjug} using the same method as for $m=2$.
Here is a sketch of the proof for $k>1$.

\smallskip
Let 
$\mathop{\tilde\Psi_k}_{N}(x)$ and $\mathop{\tilde\Phi_k}_{N}(y)$ be any
solutions of the differential systems
\be
{\partial\over \partial x}
\mathop{\tilde\Psi_k}_{N}(x) = \mathop{D_k}^N(x) \mathop{\tilde\Psi_k}_{N}(x)
\ , \qquad
{\partial\over \partial y} \mathop{\tilde\Phi_k}_{N}(y) =
 - \mathop{\tilde\Phi_k}_{N}(y) \mathop{{\tilde{D}}_k}^N(y) \ .
\ee
We construct the functions $\tilde\psi_{k,n}(x)$ with 
$ N-s_k-s_{k-1} \leq n \leq N+r_k+r_{k-1}-1$ and $\tilde\phi_{k,n}(y)$ with 
$N-r_k-r_{k-1} \leq n \leq N+s_k+s_{k-1}-1$ by recursively applying the shift 
operators (which are again compatible with the differential systems).
This gives
\be
(\partial_x + \partial_y) 
\mathop{\tilde\Phi_k}_{N}(y)  \mathop{\mathbb A_k}_N
\mathop{\tilde\Psi_k}_{N}(x) =
(x-y)
\mathop{\hat\Phi_k}_{N}(y)  \mathop{\mathbb A_{k-1}}_N
\mathop{\hat\Psi_k}_{N}(x) \ .
\ee
This equality holds term by term, since the coefficients for each monomial of
type $\tilde\psi_{k,n}(x) \tilde\phi_{k,l}(y)$ are the same as when
$\tilde\psi_{k,n}(x)=\psi_{k,n}(x)$ and $\tilde\phi_{k,l}(y)=\phi_{k,l}(y)$,
 which are linearly independent functions of $x$ and $y$.
By taking $x=y$ one has, for any $\mathop{\tilde\Phi_k}_{N}(y)$ and
$\mathop{\tilde\Psi_k}_{N}(x)$,
\be
\mathop{\tilde\Phi_k}_{N}(x)\left( \mathop{{\tilde{D}}_k}^N(x) \, 
\mathop{\mathbb A_k}_N - \mathop{\mathbb A_k}_N \, \mathop{D_k}^N(x) \right)
\mathop{\tilde\Psi_k}_{N}(x) = 0 \ .
\ee
Since this holds for any basis of solutions, the factor in brackets $(\dots)$
must vanish.
\hfill Q.E.D.

\medskip

Here is a sketch of the proof of (\ref{dualtensor}); it is very similar to the
proof of Prop. \ref{accordion}. First, we show how to prove that
\be
\det{(y\1 + D_1(x))} \propto \det{(x\1 - D_2(y))} \ ,
\ee
the other cases being similar.

First, notice that:
\bea
&&\det{(y \1 + D_1(x))}  =  {\det{\left(-\ds{{\mathop\alpha^{1
N}}_{2,-s_2}}\right)}\over \det{\left( \ds{\mathop{a_1}_{N-s_2} \dots
\mathop{a_1}_{N-1}}\right)} } \cr
&& \times \det{\left( \1  -
\mathop{a_1}_{N-s_2}{y\over \ds{{\mathop\alpha^{1 N}}_{2,-s_2}}
}\mathop{a_1}_{N-1} \dots  \mathop{a_1}_{N-s_2+1} + \sum_{l=-s_{2}+1}^{r_2}
\mathop{a_1}_{N-s_2}{\ds{{\mathop\alpha^{1 N}}_{2,l}}\over
\ds{{\mathop\alpha^{1 N}}_{2,-s_2}} } \mathop{a_1}_{N+l-1} \dots 
\mathop{a_1}_{N-s_2+1}  \right)} \ .  
\eea 
Using lemma \ref{tensor}, the last
determinant can be written as the determinant of a block matrix $T_1$ of size
$(r_1+s_1)\times (r_2+s_2)$. \be
\det{(y\1 + D_1(x))} = c_1 \, \det{(\1-T_1)} \ , \ c_1={\rm const.}
\ , \ee
where
\be
T_1:=\le[
\begin{array}{c|c|c|c}
0&\ds\mathop{a_1}_{N+r_2-1}  & 0 & 0\\ 
[7pt] \hline
0&0 & \ddots  & 0 \\
[7pt] \hline
0&0 & 0 & \ds\mathop{a_1}_{N-s_2+1} \\
[7pt] \hline
\ds{ - \mathop{a_1}_{N-s_2}{\ds{{\mathop\alpha^{1 N}}_{2,r_2}}
 \over  \ds{{\mathop\alpha^{1 N}}_{2,-s_2}} }} &  
\ds{ - \mathop{a_1}_{N-s_2}{\ds{{\mathop\alpha^{1 N}}_{2,r_2-1}}
  \over \ds{{\mathop\alpha^{1 N}}_{2,-s_2}} }}
 & 
\cdots \mathop{a_1}_{N-s_2}\ds{ { y - \ds{{\mathop\alpha^{1 N}}_{2,0}} 
 \over \ds{{\mathop\alpha^{1 N}}_{2,-s_2}} }} \cdots &
\ds{ - \mathop{a_1}_{N-s_2}{\ds{{\mathop\alpha^{1 N}}_{2,-s_2+1}} 
 \over \ds{{\mathop\alpha^{1 N}}_{2,-s_2}} }}
\end{array} \ri] \ .
\ee
On the other hand, by the same argument, we have
\be
\det{(x\1 - D_2(y))} = c_2 \, \det{(\1-T_2)} \ , \ c_2={\rm const.}
\ , \ee
where
\be
T_2:=\le[
\begin{array}{c|c|c|c}
0&\ds\mathop{a_2}_{N+r_1-1}  & 0 & 0\\ [7pt]
\hline
0&0 & \ddots  & 0 \\[7pt]
\hline
0&0 & 0 & \ds\mathop{a_2}_{N-s_1+1} \\[7pt]
\hline
\ds{ - \mathop{a_2}_{N-s_1}{\ds{{\mathop\alpha^{2 N}}_{1,r_1}}  
\over \ds{{\mathop\alpha^{2 N}}_{1,-s_1}} }} &  
\ds{ - \mathop{a_2}_{N-s_1}{\ds{{\mathop\alpha^{2 N}}_{1,r_1-1}} 
\over \ds{{\mathop\alpha^{2 N}}_{1,-s_1}} }}
 & 
\cdots \mathop{a_2}_{N-s_1}\ds{ { x - \ds{{\mathop\alpha^{2 N}}_{1,0}} 
\over \ds{{\mathop\alpha^{2 N}}_{1,-s_1}} }}
\cdots & \ds{ - \mathop{a_2}_{N-s_1}{\ds{{\mathop\alpha^{2 N}}_{1,-s_1+1}} 
\over \ds{{\mathop\alpha^{2 N}}_{1,-s_1}} }}
\end{array} \ri]  \ .
\ee
It is easy to see that $T_1$ and $T_2$ are equal up to permutations of rows
and of columns, and therefore they have the same determinant.

\smallskip

The other equalities with $k>1$,
\be
\det{\left( x \1 -  \mathop{D_{k+1}}^N(y) \right)}
\propto \det{\left( y \1 -V'_k(x) {\rm Id} +  \mathop{D_k}^N(x)
\right)} \qquad {\rm for} \  k>1 \ ,
\ee
are proved by the same method and by induction on $k$.
We define the sequence of functions $x_j(x,y)$, $1\leq j \leq k+1$, 
such that
\be
x_k=x \quad , \quad x_{k+1} = y \quad , \quad x_{j-1} = V'_j(x_j) - x_{j+1}
 \quad 2\leq j \leq k  \ .
\ee
We then prove by induction on $j$ that
\be
\det{(x_{j-1}-D_{j}(x_{j}))} \propto \det{(x_{j}-D_{j+1}(x_{j+1}))} \qquad
 2\leq j\leq k  \ .
\ee
Each step of the induction is similar to the method described above for $k=1$.
This completes the proof of \ref{dualtensor}.
\hfill Q.E.D.

\bigskip

It can also be proven that all these systems are compatible with the shifts and
deformations. It follows that if $\ds{\mathop{{\bf\Phi}_k}_N(x)}$ and 
$\ds{\mathop{{\bf\Psi}_k}_N(x)}$ denote fundamental solution matrices for
the systems  (\ref{MMdx}),  it is possible to choose their normalizations 
such that
\be
\mathop{\Phi_k}_{N}(x)  \mathop{\mathbb A_k}_N \mathop{\Psi_k}_{N}(x) =  \1 \ .
\ee
This may again be viewed as a form of the bilinear identities that allow
us to deduce bilinear equations for $\tau$-functions.

\bigskip
\noindent
{\em Acknowledgements.}  The authors would like to thank J. Hurtubise for
helpful discussions relating to this work. The first two authors (MB, BE) would
like to thank the CRM  for support throughout the period (2000-2001) in which
this work was  completed.
\bigskip

\end{document}